\documentclass{aa}    %% AASTeX
\usepackage{natbib} 
\usepackage{graphicx} 
\usepackage{amssymb}
\usepackage{epsfig}
\begin{document}
\def\sizex{16.0 cm}
\def\bigx{10.0 cm}
\def\smallerxsize{7.0 cm}
\def\smallxsize{10.0 cm}
\def\smallysize{12.0 cm}
\def\comseb#1{{\bf #1}}

\title {The Canada-France deep fields survey--II:\\ Lyman-break
  galaxies and galaxy clustering at $z\sim3$ \thanks{ Based on
    observations obtained at the Canada--France--Hawaii Telescope
    (CFHT) which is operated by the National Research Council of
    Canada, the Institut des Sciences de l'Univers (INSU) of the
    Centre National de la Recherche Scientifique and the University of
    Hawaii, and at the Cerro Tololo Inter--American Observatory and
    Mayall 4-meter Telescopes, divisions of the National Optical
    Astronomy Observatories, which are operated by the Association of
    Universities for Research in Astronomy, Inc.  under cooperative
    agreement with the National Science Foundation.}}
\titlerunning{Lyman-break galaxies in the CFDF survey}
\authorrunning{S. Foucaud et al.}
 
\author{ S.\ Foucaud \inst{1,9} \and H.\ J.\ McCracken \inst{1,8} \and
  O.  Le F\`evre \inst{1} \and S.\ Arnouts \inst{2,1} \and M.\ Brodwin
  \inst{3} \and \\ S.\ J.\ Lilly \inst{4} \and D. Crampton \inst {5}
  \and Y.\ Mellier \inst{6,7}} \institute{Laboratoire d'Astrophysique
  de Marseille, Traverse du Siphon, 13376 Marseille Cedex 12, France
  \and ESO - European Southern Observatory, Karl-Schwarzschild-Str. 2,
  85748 Garching bei M\"unchen, Germany \and University of Toronto,
  Department of Astronomy, 60 St. George Street, Toronto, Ontario,
  Canada M5S 3H8 \and Institute of Astronomy - ETH Hoenggerberg, HPF
  D8, 8093 Zurich, Switzerland \and Herzberg Institute for
  Astrophysics, 5071 West Saanich Road, Victoria, British Colombia,
  Canada V9E 2E7 \and Institut d'Astrophysique de Paris, 98 bis
  Boulevard Arago, 75014 Paris, France \and Observatoire de Paris,
  LERMA, 61 Avenue de l'Observatoire, 75014 Paris, France \and Present
  address: University of Bologna, Department of Astronomy, via Ranzani
  1, 40127 Bologna, Italy \and Present address: Istituto di
  Astrofisica Spaziale e Fisica cosmica - Sezione di Milano, via
  Bassini 15, 20133 Milano, Italy}
\offprints{S. Foucaud,\\ e-mail: \texttt{foucaud@mi.iasf.cnr.it}}

\abstract{We present a large sample of $z\sim3$ $U-$ band dropout
  galaxies extracted from the Canada-France deep fields survey (CFDF).
  Our catalogue covers an effective area of $\sim 1700$ arcmin$^2$
  divided between three large, contiguous fields separated widely on
  the sky. To $I_{AB}=24.5$, the survey contains 1294 Lyman-break
  candidates, in agreement with previous measurements by other
  authors, after appropriate incompleteness corrections have been
  applied to our data. Based on comparisons with spectroscopic
  observations and simulations, we estimate that our sample of
  Lyman-break galaxies is contaminated by stars and interlopers
  (lower-redshift galaxies) at no more than $\sim 30\%$.  We find that
  $\omega(\theta)$ is well fitted by a power-law of fixed slope,
  $\gamma=1.8$, even at small ($\theta<10''$) angular separations.  In
  two of our three fields, we are able to fit simultaneously for both
  the slope and amplitude and find $\gamma = 1.8 \pm 0.2 $ and $r_0 =
  (5.3^{+6.8}_{-2.2})h^{-1}$~Mpc, and $\gamma = 1.8 \pm 0.3 $ and $r_0
  = (6.3^{+17.9}_{-2.8})h^{-1}$~Mpc (all spatially dependent
  quantities are quoted for a $\Lambda$-flat cosmology).  Our data
  marginally indicates in one field (at a $3 \sigma$ level) that the
  Lyman-break correlation length $r_0$ depends on sample limiting
  magnitude: brighter Lyman-break galaxies are more clustered than
  fainter ones. For the \textit{entire} CFDF sample, assuming a fixed
  slope $\gamma=1.8$ we find $r_0=(5.9\pm0.5)h^{-1}$~Mpc.  Using these
  clustering measurements and prediction for the dark matter density
  field computed assuming cluster-normalised linear theory, we derive
  a linear bias of $b=3.5\pm0.3$.  Finally we show that the dependence
  of the correlation length with the surface density of Lyman-break
  galaxies is in good agreement with a simple picture where more
  luminous galaxies are hosted by more massive dark matter halos with
  a simple one-to-one correspondence.  
  }

\date{received ... ; accepted ... }
\maketitle

\section{Introduction}
\label{sec:int}

Surveys of the local Universe, such as the Two-degree Field Galaxy
Redshift Survey (2dFGRS) \citep{2001MNRAS.328.1039C} and the Sloan
Digital Sky Survey (SDSS) \citep{2002AJ....123..485S}, are now
providing ever-more detailed pictures of the distribution of galaxies
on scales of several hundred Mpc.  We have a good paradigm of
large-scale structure formation in which small fluctuations of matter
density grow under the influence of gravity to form large-scale
structures and galaxy halos. Furthermore perturbation theory and
numerical simulations provide useful predictions which can be
challenged against observations.  Making some assumptions on how dark
matter traces luminous objects at large scales, we can produce a
picture of how galaxies are distributed on large scales locally which
match observational data remarkably well.

However, predicting the evolution of clustering to higher redshift is
still challenging. We may either attempt to construct a fully
self-consistent model of galaxy formation which links the dark matter
distribution and the luminous galaxies (e.g.
\citealp{1994MNRAS.271..781C}; \citealp{1999MNRAS.305L..21B}) or,
alternatively, to postulate a relationship between dark matter halos
and luminous galaxies (the bias) and use this to predict the galaxy
distribution (e.g.  \citealp{1997MNRAS.286..115M};
\citealp{1999MNRAS.304..175M}). One simple version of this method has
been to postulate a linear relationship between the galaxy density,
$\delta_0$, and the dark matter one, $\delta_m$: $b=\delta_0/\delta_m$
, where $b$ is the bias parameter \citep{1984ApJ...284L...9K}.
However, until recently, comparing these models to available
observations has not been straightforward. For example, angular
clustering analyses (e.g.  \citealp{RSMF};
\citealp{2000MNRAS.318..913M}) are usually based on magnitude limited
samples that typically contain a mixture of galaxy types within a range
of redshifts and thus require additional information on the evolution
of the galaxy population to allow us to draw meaningful conclusions
about the evolution of galaxy clustering.

A much more powerful technique is to measure the clustering of
galaxies isolated in different redshift intervals. Spectroscopic
surveys, such as the Canada-France Redshift Survey (CFRS)
\citep{1995ApJ...455...50L,1996ApJ...461..534L} and the Canadian
Network for Observational Cosmology survey (CNOC)
\citep{2000ApJ...542...57C}, allow us to directly measure the
evolution of the galaxy correlation length $r_0$ as a function of
redshift. Alternatively, the photometric redshift technique has
enabled similar analyses up to fainter magnitudes and higher redshifts
(e.g.
\citealp{1999MNRAS.310..540A,2000ApJ...541..527B,2002MNRAS.329..355A}).
Although these various samples are subject to different selection
effects and cosmic variance, the results on the clustering
measurements agree in showing a general decline of the comoving
correlation length $r_0$ with redshift from $z\sim 0$ to $z\sim 1$.
While the clustering amplitude of the underlying dark matter is also
expected to decrease with look-back time with a rate depending on the
cosmological parameters, the above observations cover a too small
redshift range to provide constraints on the evolution of galaxy
clustering.

In the early 1990's, several studies attempted to photometrically
isolate high redshift ($z\sim3$) galaxies using very deep $U-$ band
imaging \citep{1990ApJ...357L...9G,1993AJ....105.2017S}. The Lyman
limit discontinuity in the emission light of these (star-forming)
galaxies, combined with absorption by the intergalactic medium along
the line of sight \citep{1995ApJ...441...18M,1999ApJ...518..103B}
means these objects are expected to have extremely red $(U-B)$
colours, and $(V-I)$ colours about zero \citep{1996MNRAS.283.1388M}.
However, it was the advent of 10-m telescopes which allowed the
redshifts of these galaxies to be spectroscopically confirmed
\citep{1996ApJ...462L..17S}.  Today, a thousand or so of these bright
galaxies (i.e. those with $L\sim L^{*}$) have been spectroscopically
confirmed at redshift $z\sim 3$ (Lyman-break galaxies --
\citeauthor{1999ApJ...519....1S},~\citeyear{1999ApJ...519....1S}),
whereas previously only peculiar objects such as QSOs or
radio-galaxies were known at this epoch.

The most recent works on $z\sim3$ galaxies have focused on their
physical properties: for example, \cite{2003ApJ...584...45A}
investigated the cross-correlation between Lyman-break galaxies and
the intergalactic medium whereas \cite{2003ApJ...588...65S} studied
their rest-frame UV spectroscopic properties. Properties of these
objects at other wavelengths have also been investigated
\citep{2002ApJ...576..625N, webb2003}.
  
More recently, the focus has shifted to replicating the selection of
high-redshift objects using the drop-out technique at $z\sim4$ and
beyond
\citep{1999ApJ...519....1S,2001MNRAS.325..897S,2001ApJ...558L..83O,lehnert2002}.

Early studies of clustering measurements of Lyman-break galaxies
selected photometrically \citep{1998ApJ...503..543G} and
spectroscopically \citep{1998ApJ...505...18A} indicated they have a
correlation length of $\sim 4 h^{-1}$~Mpc, comparable to nearby
massive galaxies ; a result confirmed by more recent studies (e.g.
\citeauthor{2003ApJ...584...45A},~\citeyear{2003ApJ...584...45A}).
Since the strength of clustering for dark matter is expected to
continuously decrease towards higher redshifts, the high clustering
amplitudes found at $z\sim 3$ implies that Lyman-break galaxies are
biased tracers of the underlying dark matter distribution, and
furthermore suggests that they form preferentially in massive dark
matter halos.  In the current theoretical paradigm, more massive
objects, which form at rarer peaks in the underlying dark matter
distribution, have clustering amplitudes much higher than those of
less massive, less luminous galaxies
\citep{1984ApJ...284L...9K,1986ApJ...304...15B}. More recent analyses
of these Lyman-break galaxies datasets focused on the dependence of
clustering amplitude on apparent magnitude selection
\citep{2001ApJ...550..177G} or the behaviour of the galaxy clustering
at small angular scales \citep{2002ApJ...565...24P}. However, the
angular scales probed are generally small, as these surveys consist of
many non-contiguous fields each of which covers $\sim 50$~arcmin$^2$.
These samples generally contained too few objects to allow a reliable
detection of clustering dependence on apparent magnitude or to place
useful constraints on the slope of the galaxy correlation function.
Furthermore there is also a large spread of measurements made at the
same limiting magnitude suggesting the presence of systematic effects
or cosmic variance in these surveys.
 
In this paper, the second in a series, we report new measurements of
number counts and clustering properties of Lyman-break galaxies
selected in the Canada-France Deep Fields survey (CFDF). The CFDF is a
deep, wide-field multi-wavelength survey of four unconnected fields
covering three of the CFRS fields.  In \citet{cfdf1}, hereafter
referred as Paper I, we described the global properties of the CFDF
sample and presented measurements of the two-point galaxy correlation
function $\omega(\theta)$ as a function of angular scale, limiting
$I_{AB}$ magnitude and $(V-I)_{AB}$ colour.

Our wide field optical imaging, combined with deep $U-$ band imaging,
covering $\sim0.65$~deg$^2$, allows us to construct the largest sample
of photometrically selected Lyman-break galaxies to date. Using
spectroscopic observations and simulated catalogues we demonstrate our
selection criterion is robust and estimate the degree of contamination
in our catalogues. Our three fields, each covering scales of
$28\arcmin$ and separated widely on the sky, allow us to make a robust
estimation of the effect of cosmic variance on our results.
Additionally the large angular scale of each CFDF field allows us to
probe comoving separation at least twice larger than previous works
($\sim9h^{-1}$~Mpc at redshift $z\sim3$ for a $\Lambda$-flat cosmology
with $\Omega_0=0.3, \Omega_{\Lambda}=0.7$).

This paper is organised as follows: in Section~\ref{sec:observ-reduct}
we briefly describe the observations which comprise the CFDF survey;
in Section~\ref{sec:select-lyman-break} we outline how Lyman-break
galaxies were selected in the CFDF, and present an estimate of the
robustness of this selection; in Section~\ref{sec:clust_LBG} we
present our clustering measurements of the CFDF Lyman-break sample; in
Section~\ref{sec:comp-with-theor} we compare these observations to a
range of theoretical predictions, and present our interpretation.
Finally, conclusions are summarised in
Section~\ref{sec:summary-conclusions}. Unless stated otherwise,
throughout this paper we use a $\Lambda$-flat cosmology
($\Omega_0=0.3$, $\Omega_{\Lambda}=0.7$ to compute spatial quantities
and we assume $h=H_0/100$ km.s$^{-1}$.Mpc$^{-1}$).

\section{Observations, data reductions and catalogue preparation}
\label{sec:observ-reduct}

\subsection{Observations and data reductions}
\label{sec:observ-data-reduct}

The CFDF survey comprises four separate $28\arcmin\times28\arcmin$
fields; and for two and half of these fields we have complete $UBVI$
photometry. In total these fields cover $\sim0.65$ deg$^2$ and include
the 03hr, 14hr and 22hr fields of the CFRS survey
\citep{1995ApJ...455...50L}.  Lyman-break studies have already been
carried out in several subareas of the CFDF-14hr (the ``Groth strip'')
and the CFDF-22hr fields by \citet{1999ApJ...519....1S}.

Full details of the CFDF $BVI$ observations and the data reduction
procedures are given in Paper I.  These observations were carried out
using the University of Hawaii's 64-megapixel mosaic camera (UH8K) at
the Canada-France Hawaii Telescope (CFHT) in a series of runs from 1996
to 1997. In Paper I we demonstrated that the $I_{AB}$ zero-point r.m.s
magnitude variation across each UH8K pointing is $\sim 0.04$
magnitudes, and that our internal r.m.s. astrometric accuracy (between
images taken in separate filters) is $\sim 0.05''$.  This allows us to
measure accurately galaxy colours by using the same aperture at the
same $(x,y)$ position on stacks constructed from different filters
without the needing to positionally match our catalogues.

The unthinned Loral-3 CCDs used in UH8K has very poor response
blueward of $4000$\AA. For this reason, separate $U-$ band
observations were carried out at the Cerro Tolo Inter-American
Observatory (CTIO) and at the Kitt Peak National Observatory (KPNO)
4-m telescopes. The detectors used were TEK $2048\times2048$ thinned
CCDs with a pixel scale of $0.42''$~pixel$^{-1}$. To cover each
$28\arcmin\times28\arcmin$ UH8K field, four separate pointings were
required. Total integration per pointing was approximatively $10$
hours with 10 to 15 exposures. Within each pointing, the airmass
varied between 1.0 and 1.6 and seeing ranged from 1.0$''$ to 1.4$''$.
Reduction of these data followed the usual steps of bias and overscan
removal followed by flat-fielding.  Each exposure in each pointing was
then stacked and scaled so that all have the same photometric
zero-point. A coordinate transformation between each of the four
sub-pointings and the CFDF $I-$ band was then computed.  These
sub-pointings were then resampled using this transformation to the
pixel scale of UH8K ($0.205''$pixel$^{-1}$).  Finally, each
sub-pointing was coadded to make a single large mosaic covering the
entire field of UH8K. The 14hr and 03hr fields consist of four
separate $U-$ band sub-pointings, whereas we have only two for the
22hr data.

\begin{table}
\begin{tabular}{*{5}{c}}
{\bfseries Field}   & R.A.     & Dec.      & Band & $3\sigma$ limit\\
                    & (J2000)   & (J2000)    &      & ($AB$ mags) \\

\hline                                                                       
                    &          &           &   &       \\
{\bfseries 0300+00} & 03:02:40 & +00:10:21 & U & 26.98 \\
                    &          &           & B & 26.38 \\
                    &          &           & V & 26.40 \\
                    &          &           & I & 25.62 \\
                    &          &           &   &       \\
{\bfseries 1415+52} & 14:17:54 & +52:30:31 & U & 27.71 \\
                    &          &           & B & 26.23 \\
                    &          &           & V & 25.98 \\
                    &          &           & I & 25.16 \\
                    &          &           &   &       \\
{\bfseries 2215+00} & 22:17:48 & +00:17:13 & U & 27.16 \\
                    &          &           & B & 25.76 \\
                    &          &           & V & 26.18 \\
                    &          &           & I & 25.22 \\
                    &          &           &   &       \\
\end{tabular}
\caption{Details of the CFDF images used in this study. For the 03hr and 14hr
fields, we list the $3\sigma$ detection  limit inside an aperture of
$3''$ for images convolved to the worst seeing (i.e. 1.3$''$ for 03hr
and 1.4$''$ for 14hr). For the 22hr field, we list the $3\sigma$ detection limit  
inside an aperture of $4''$ for un-convolved images.}

\label{tab:cfdf.fields}
\end{table}

\subsection{Catalogue preparation}
\label{sec:catal-prep}

As described in Paper I, we prepared catalogues using the $\chi^2$
technique outlined in \citet{1999AJ....117...68S}. This method
provides an optimal way for detecting faint objects in multi-colour
space. We did not use this method for our 22hr data as the seeing
differs greatly across the $U-$ band images; 22hr $U-$ band images are
composed of two different pointings taken at CTIO, one has a seeing of
1.2$''$ and the second 1.4$''$. Application of the $\chi^2$ technique
would involve convolving all images in all bands to the worst seeing,
which we would prefer to avoid. Instead, we use an object detection
list generated from the $I-$ band image and measure colours using
apertures at these positions for the other four images.  To account
for the poorer seeing in these images we use a slightly larger
aperture of $4''$ to measure galaxy colours; for the other fields we
use an aperture of $3''$. As we will see later, the slightly different
reduction procedures used for the 22hr field does not affect our
clustering measurements. Throughout, galaxy magnitudes are measured
using \cite{1980ApJS...43..305K} total magnitudes (\texttt{SExtractor}
parameter \texttt{MAG\_AUTO}, \citealp{1996A&AS..117..393B}).
Table~\ref{tab:cfdf.fields} gives the central coordinates of the three
fields and the limiting magnitudes in the different bands, taking into
account the different aperture sizes and extraction methods used to
prepare each catalogue.

Polygonal masks were created covering regions near bright stars, or
with lower signal-to-noise, and objects inside these areas were
rejected. The total area, after masking, is given in
Table~\ref{tab:samplim}. As explained in Paper I (section 5.1), we
have conducted extensive tests with both correlated and uncorrelated
mock datasets to demonstrate that the masking procedure does not
affect the estimated correlation amplitudes.

\section{The sample of Lyman-break galaxies}
\label{sec:select-lyman-break}

\subsection{Selecting Lyman-break galaxy candidates}
\label{sec:select-lyman-break-1}

\begin{figure}
\begin{center}
 \vspace{0.2cm}\resizebox{7.5cm}{!}{\includegraphics{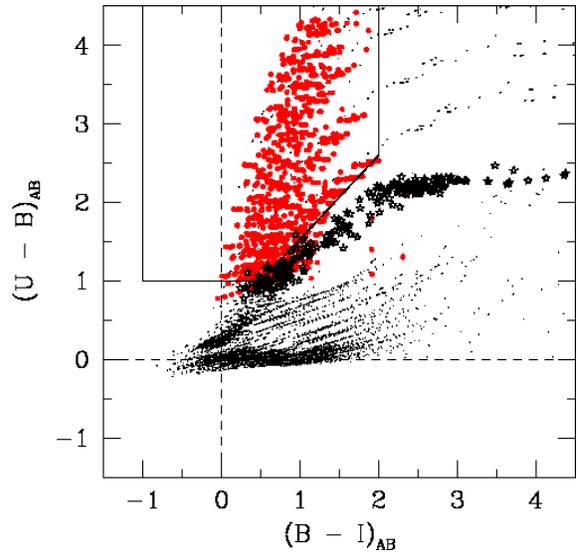}}\vspace{0.8cm}
\caption{Galaxy evolutionary tracks (dots) used to define our selection box,
  represented as the solid line.  Filled symbols indicate galaxies in
  the range $2.9<z<3.5$. Star symbols represent simulated colours for
  galactic stars with $I_{AB}<20.0$.}
\label{fig:selsim}
\end{center}
\end{figure}

\begin{figure}
\begin{center}
  \vspace{0.2cm}\resizebox{7.5cm}{!}{\includegraphics{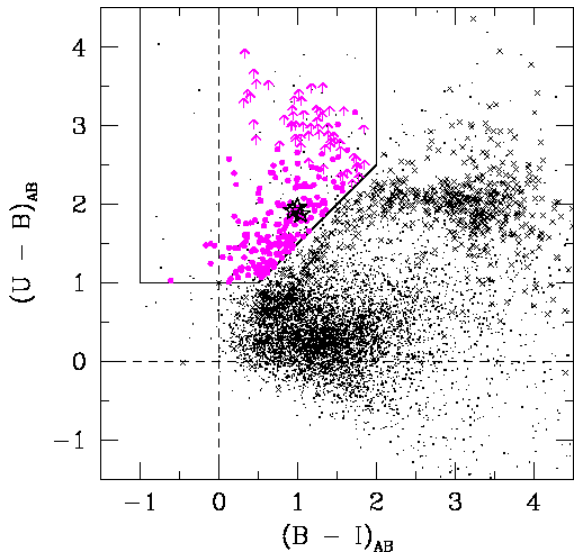}}\vspace{0.8cm}
\caption{$(U-B)_{AB}$ against $(B-I)_{AB}$ for galaxies with $I_{AB}<24$ in
  the CFDF-03hr field. Almost 14,000 objects are represented; for
  clarity only half of all objects in this field are shown. The solid
  line represents the selection box given in Equation~\ref{eq:selbox}.
  There are 269 candidates (filled circles) which satisfy our selection
  criteria. The arrows indicate Lyman-break candidates which have a
  $1\sigma$ upper limit in $U$. Crosses indicate star-like objects
  (identified on the basis of their compactness).  The three
  spectroscopically confirmed Lyman-break galaxies are shown with star
  symbols. }
\label{fig:03hrsel}
\end{center}
\end{figure}

Lyman-break candidates were selected by isolating the Lyman-break
feature at $912$\AA\ in a colour-colour diagram
\citep{1993AJ....105.2017S}. To define our selection box we examine
the path of synthetic evolutionary tracks in the $(U-B)$ vs $(B-I)$
colour-colour space defined by the CFDF filter set. These tracks are
derived from a set of spectral energy distribution templates
\citep{BC} assuming a $\Lambda$-flat cosmology.

Figure~\ref{fig:selsim} illustrates the tracks used; each track
represents a different combination of galaxy type, age, metallicities
and reddening.  Internal extinction is modelled using a relation
appropriate for starburst galaxies \citep{1994ApJ...429..582C}. We
have also included the Lyman absorption produced by the intergalactic
medium following \citet{1995ApJ...441...18M}.  Colours of field stars
are estimated using the galactic model of \citet{1986A&A...157...71R}
transformed to our instrumental system (magnitude errors are not
include in this Figure).

Based on these considerations, we define our selection box as
 
\begin{equation}
\begin{array}{rcccl}
1.0&\leq&(U-B)_{AB},&&\\
-1.0&\leq&(B-I)_{AB}&\leq&2.0,\\
(B-I)_{AB}+0.5&\leq&(U-B)_{AB},&&\\
&&(V-I)_{AB}&\leq&1.0.\\
\end{array}
\label{eq:selbox}
\end{equation}

We estimate the redshift range of our Lyman-break sample to be
$2.9<z<3.5$, quite close to the $2.7<z<3.4$ interval sampled by
\citet{1996AJ....112..352S}.

The criterion $(V-I)_{AB}\leq1.0$ reduces contamination by stars and
avoids contamination by elliptical galaxies $z\sim1.5$. Additionally,
we require that our Lyman-break candidates are detected in $B$, $V$ and
$I$. Finally, \textit{all} candidates are visually inspected in all
five channels ($UBVI$ and the $\chi^2$ detection image) before they are
added to the source catalogue. About ten percent of the Lyman-break
sources were rejected as spurious; these objects are typically
detections on bad columns or other cosmetic defects which had not been
removed by the masking process. Given the detection limits in $U$ and
$I$ presented in Table~\ref{tab:cfdf.fields}, selecting Lyman-break
galaxies to $I_{AB}=24.5$ is feasible for all our fields.

\begin{figure}
  \resizebox{\hsize}{!}{\includegraphics{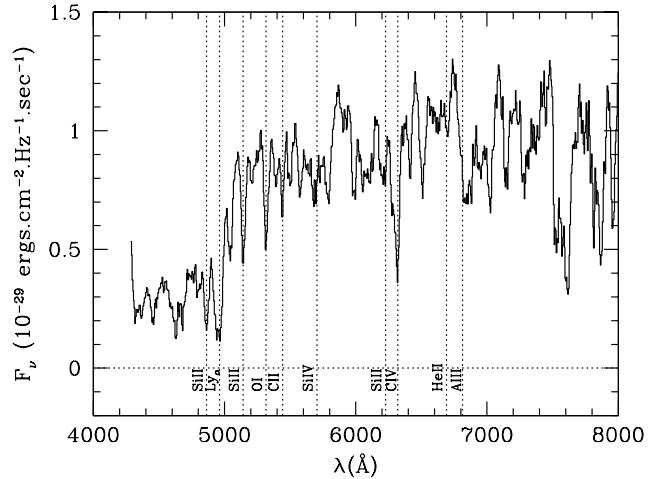}}
\caption{Spectrum of a confirmed Lyman-break galaxy at
  redshift $z=3.08$ observed at CFHT using the MOS spectrograph.
  Spectral features are indicated with the dotted lines.}
\label{fig:spec}
\end{figure}

In Figure~\ref{fig:03hrsel} we show the $(U-B)$ vs $(B-I)$
colour-colour diagram for galaxies with $I_{AB}<24$ in the 03hr field,
with Lyman-break candidates identified using the selection box defined
in Equation~\ref{eq:selbox}. Redshifts for three of these galaxies
were spectroscopically confirmed ($z=3.07,3.08$ and $3.27$
respectively) with data taken at CFHT in November 1997 using the
Multi-Object Spectrograph. These galaxies are plotted in
Figure~\ref{fig:03hrsel} as open stars. A spectrum of one of these
galaxies is shown in Figure~\ref{fig:spec}.

As we have limited spectroscopy on CFDF $z\sim3$ galaxies, we have
carried out extensive simulations, described in the
Section~\ref{sec:robustn-select-box}, to ensure the robustness of our
selection box. As we will see, these simulations allow us to quantify
how much contamination we expect from stars and lower-redshift
interlopers.

\subsection{Source counts of Lyman-break galaxies}
\label{sec:sources-counts}

\begin{figure}
  \resizebox{\hsize}{!}{\includegraphics{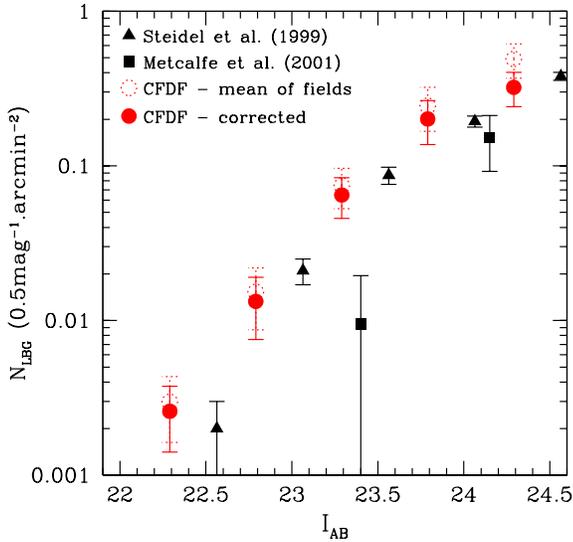}}
\caption{Raw and corrected number counts of Lyman-break galaxies in the CFDF
  (open and filled circles respectively). The errorbars on each point
  is computed from the amplitude of the field-to-field variance.  We
  also show colour-selected Lyman-break galaxy counts from
  \citet{2001MNRAS.323..795M} (filled squares) and
  \citet{1999ApJ...519....1S} (filled triangles).}
\label{fig:ngalcounts}
\end{figure}

In Figure~\ref{fig:ngalcounts} we present our raw and corrected
Lyman-break galaxy counts as a function of $I_{AB}$ magnitude (dotted
and solid symbols).  Table~\ref{tab:samplim} presents the surface
densities of our Lyman-break galaxy sample for a ranges of limiting
magnitudes. In addition, we present counts from
\cite{2001MNRAS.323..795M} and \cite{1999ApJ...519....1S} samples of
Lyman-break galaxies. The latter compilation contains a correction for
contamination by stars and AGN estimated from spectroscopic
observations.

To convert these $\mathcal{R}_{AB}$-selected observations to our
$I_{AB}$ magnitudes, we estimate the mean colour of Lyman-break
galaxies at redshift $z\simeq3$ to be $(R-I)_{AB}\simeq0.3$. In making
this transformation we assume that the colours of the Lyman-break
population do not evolve with magnitude. Combining this with the
conversion between $\mathcal{R}_{AB}$ and $R_{AB}$ given in
\cite{1993AJ....105.2017S}, we estimate that
$(\mathcal{R}-I)_{AB}\simeq0.19$.
  
At bright magnitudes, our counts are in good agreement with the
literature compilation: however, at fainter bins $24.0<I_{AB}<24.5$
they exceed the literature comparisons by a factor $1.3-1.5$.
Essentially this is due to higher contamination in our sample.
According to simulations (which we describe fully in
Section~\ref{sec:robustn-select-box}) this contamination, arising from
the shallower depth of our $UBVI$ data compared to
\citeauthor{1999ApJ...519....1S}'s $U_n G \mathcal{R}$ data, amounts
to $\sim 30\%$ in the faintest bins. Counts corrected for this
contamination are indicated as the dotted symbols in
Figure~\ref{fig:ngalcounts}.  After this correction our counts are in
closer agreement with the literature.

\begin{table*}[htbp]
\begin{center}
\begin{tabular}{{c}*{3}{cp{1cm}}c}
 & \multicolumn{2}{c}{{\bfseries 0300+00}} &
 \multicolumn{2}{c}{{\bfseries 1415+52}} &
 \multicolumn{2}{c}{{\bfseries 2215+00}} & {\bfseries CFDF} \\
 & \multicolumn{2}{c}{{\bfseries\scriptsize{A=646$\,$arcmin$^2$}}}
 & \multicolumn{2}{c}{{\bfseries\scriptsize{A=708$\,$arcmin$^2$}}}
 & \multicolumn{2}{c}{{\bfseries\scriptsize{A=317$\,$arcmin$^2$}}} &\\
\hline
magnitude range & n & (N) & n & (N) & n & (N) & n \\
($I_{AB}$) & & & & & & & \\
\hline
 & & & & \\
20.0--22.5 & 0.003$\pm$0.003 & (2) & 0.010$\pm$0.004 & (7) & 0.006$\pm$0.006 & (2) & 0.006$\pm$0.004 
\\
22.5--23.0 & 0.020$\pm$0.006 & (12) & 0.008$\pm$0.004 & (6) & 0.013$\pm$0.008 & (4)  & 
0.014$\pm$0.006 \\
23.0--23.5 & 0.08$\pm$0.01 & (55) & 0.05$\pm$0.01 & (36) & 0.07$\pm$0.02 & (22)  & 0.07$\pm$0.02 \\
23.5--24.0 & 0.31$\pm$0.02 & (200) & 0.16$\pm$0.02 & (116) & 0.22$\pm$0.03 & (69)  & 0.23$\pm$0.08 
\\
24.0--24.5 & 0.59$\pm$0.03 & (379) & 0.34$\pm$0.02 & (242) & 0.45$\pm$0.04 & (142)  & 0.46$\pm$0.13 
\\
 & & & & \\
\end{tabular}
\caption{Differential number counts, $N$, and surface density, $n$ (in
  arcmin$^{-2}$), of Lyman-break galaxies in the CFDF fields, for a
  range of $I_{AB}$-selected slices. The mean surface density,
  labelled CFDF, is also given.  Errors in the surface density
  measurements for each individual field are computed using Poisson
  counting statistics; field--to--field variance is used to estimate
  the error in the mean.}
\label{tab:samplim}
\end{center}
\end{table*}

We note that the dispersion in Lyman-break
counts between our three fields is larger than one would expect based
on purely Poissonian errors. We suggest several possible explanations
for this dispersion. Firstly, Lyman-break galaxies are strongly
clustered objects: at the magnitude limit of the survey, this
clustering can produce count fluctuations of $\sim 15\%$. Secondly, the
absolute photometric calibration between each of the three fields
(which were all taken in different observing runs, in different
seasons, and in some cases with different $U-$ band imagers) differs by
at worst $\sim0.1$~magnitudes (although, as demonstrated in Figure~9 of
Paper I, the field--to--field variation in galaxy counts is still very
small).

How large an effect could a systematic error of $\sim0.1$~magnitude
have on the Lyman-break number counts? To address this question we
have carried out a set of simulations in which a small, Gaussian error
of $\sigma=0.1$ is added to each filter, i.e., new magnitudes are
computed according to $M'=M+\delta M$. The number counts of galaxies
falling in the selection box is recomputed at each iteration.  From
this experiment we find that magnitude errors of $\sigma=0.1$ can
produce a fluctuation in Lyman-break counts of $\sim 15\%$. Adding the
contribution from the clustered nature of Lyman-break galaxies, this
leads to a total expected field-to-field fluctuation of $\sim 20\%$,
large enough to explain the deviation between the 14hr and 22hr
fields.  We have examined the 03hr field in more detail, and we find
that one quadrant has a $\sim 50\%$ higher density of Lyman-break
candidates than the other three: if this quadrant is removed, the
fluctuations between the 03hr field and the other two can be explained
by the sources of errors listed above. The effect of this over-dense
quadrant on $\omega(\theta)$ is to increase the amount of power at
$\sim 0.1^{\circ}$ scales but, as we will see in
Section~\ref{sec:clust_LBG}, at the scales we normally measure galaxy
correlation amplitudes, the field-to-field variation in
$\omega(\theta)$ is still less than the amplitude of the Poissonian
error in $\omega(\theta)$.

\subsection{Estimating the reliability of the CFDF Lyman-break
  selection box}
\label{sec:robustn-select-box}

\begin{table*}[ht]
\begin{center}
\begin{tabular}{*{7}{c}}
magnitude range & \multicolumn{2}{c}{density of objects} & LBG & 
\multicolumn{3}{c}{contamination from} \\
($I_{AB}$)  & \multicolumn{2}{c}{in the selection box from} &  found in the &
 interlopers outside & interlopers & stars \\
 & observations & simulations & selection box & our $2.9<z<3.5$ range & with $z<2$ & \\
\hline               
 &  &  &  &  &  & \\
20.0--23.5 & 0.069 & 0.073 & 91.8\% &  7.7\% &  2.2\% & 5.5\%\\
20.0--24.0 & 0.233 & 0.233 & 89.7\% & 11.9\% &  6.3\% & 4.3\%\\
20.0--24.5 & 0.574 & 0.652 & 83.9\% & 24.5\% & 17.3\% & 3.5\%\\
 &  &  &  &  &  & \\
23.5--24.0 & 0.164 & 0.160 & 88.7\% & 13.9\% &  8.2\% & 3.8\%\\
24.0--24.5 & 0.341 & 0.419 & 80.2\% & 31.4\% & 23.5\% & 3.1\%\\
 &  &  &  &  &  & \\
\end{tabular}
\caption{Observed and simulated surface densities of objects (in
  arcmin$^{-2}$) recovered using the selection box
  (Equation~\ref{eq:selbox}), based on simulations described in
  Section~\ref{sec:robustn-select-box} and observations in the
  CFDF-14hr field. We also estimate the fraction of Lyman-break
  galaxies (LBG) recovered using this selection box from the total
  galaxy population in this redshift range.  Additionally, we present
  the fraction of contaminants within this selection box by
  interlopers (lower-redshift galaxies) outside our redshift range
  ($2.9<z<3.4$) and by interlopers with $z<2$ and by stars.}
\label{tab:contam}
\end{center}
\end{table*}

To estimate the level of contamination by stars and interlopers
(lower-redshift galaxies) and the fraction of Lyman-break galaxies
which could be missed in our sample, we construct multi-colour mock
catalogues which incorporate all the observational uncertainties.

We use the model\footnote{http://www.obs-besancon.fr/www/modele/} of
\citet{1986A&A...157...71R} to generate our stellar catalogue at the
galactic latitude of the 14hr field. The catalogue's $UBVI$
Johnson-Cousins colours were transformed to our instrumental system
and then convolved with a function describing the dependence of
magnitude errors with magnitude for each passband. In
Figure~\ref{fig:selsim} star symbols show objects from this catalogue;
for clarity, only stars with $I_{AB}<20.0$ and without magnitude
errors are shown. Fainter objects occupy the same region in
colour-colour space.

For the galaxy catalogues, we use the empirical approach developed by
Arnouts et al. (2003, in preparation); the main components of which
are as follows.  To characterise the spectral energy distribution
(SEDs) of galaxies, we use the four observed SEDs of
\citet{1980ApJS...43..393C} (corresponding to Elliptical, Sbc, Scd and
Irregular local galaxy types), and two SEDs corresponding to
star-forming galaxies with ages of 0.05 and 2 Gyrs. These SEDs were
computed using the GISSEL model \citep{BC} assuming solar metallicity,
a Salpeter initial mass function and constant star formation rate.
Following the approach adopted by \citet{1997AJ....113....1S}, we
interpolated between the 6 original spectra to provide a finer grid of
the spectral-type coverage producing a total number of 61 templates.
We derive the density of objects for given magnitude and redshift
interval using the luminosity function parameters from the $R-$ band
ESO-Sculptor Survey to $z \simeq 0.6$. Galaxies are divided into three
spectral classes: early, intermediate and late types (de Lapparent et
al. 2002, in preparation).  At higher redshift the luminosity function
parameters have been adjusted in order to reproduce the observed
redshift distributions of the CFRS \citep{1995ApJ...455...96C} and the
North and South Hubble Deep Fields (HDF-N and -S)
\citep{1999MNRAS.310..540A,2002MNRAS.329..355A}.  We derive magnitudes
in other passbands using these SEDs to compute the appropriate
$k-$correction.  A model for the ``observed'' magnitudes is obtained
by taking into account the luminosity profile of the galaxy and
observational conditions (such as seeing and surface brightness
limits) and computing the fraction of light lost according to the
magnitude scheme employed.  We derive photometric errors using the
observed dependence of error with magnitude in each passband.  This
empirical method reproduces the main observables such as counts,
colours and redshift distributions.  Special attention is paid to the
redshift distributions to ensure a reasonable description of the
relative fraction of galaxies at low and high redshift which is the
first step in quantifying how target selection in a colour-colour
diagram can be subject to contamination effects.

In Table~\ref{tab:contam} we show the surface densities of all the
simulated objects (computed for an area of $\sim 1$~deg$^2$) and
compare them to observations in the CFDF-14hr field.  The total surface
densities of objects found in the selection box from simulations and
observations are close, reflecting our requirement that the models
match observed redshift distributions.  According to the simulations,
the contamination by stars decreases from $6\%$ to $3\%$, while the
contamination by galaxies outside our chosen redshift range increases
from $8\%$ to $25\%$ for $I_{AB}<23.5$ to $I_{AB}<24.5$ respectively.
Furthermore we find that the class of interlopers changes as a function
of limiting magnitude.  For $I_{AB}<23.5$, about $70\%$ of the galaxy
interlopers are expected to be at $z\ge 2$ and the remaining at $z< 2$.
At $I_{AB}<24.5$ the situation is different, due to the larger
uncertainties in the colours: about $60\%$ of interlopers are expected
to be $z< 2$ and a large part of the remainder ($\sim 25\%$) are at
$z\ge 2$. In the following section we assess the reliability of these
simulations by direct comparison with spectroscopic observations.

Our 14hr field covers the ``Groth strip'' field.  C.~Steidel has
kindly provided us with spectroscopic redshifts for 335
photometrically selected objects in this area and we have used this
dataset to assess the reliability of our selection box. There are 315
objects in common (based on a simple positional match) between the two
catalogues, and for these objects, selected using $U_n G \mathcal{R}$
photometry, we have spectroscopic redshifts in addition to CFDF $UBVI$
photometry.  Table~\ref{tab:cfdfstei} shows the redshift distribution
for galaxies with $I_{AB}<24.5$ before and after the application of
our selection box.
\begin{table}[htbp]
\begin{center}
\begin{tabular}{*{3}{c}}
redshift      & Steidel et al. & combined CFDF/Steidel \\
range         & box            & box                   \\
\hline
total         &  108           & 52                    \\
$2.9<z<3.5$   &  52            & 31                    \\
$2.0<z<2.9$   &  26            & 13                    \\
$z \leq 2.0$  &  0             & 0                     \\
$z \geq 3.5$  &  1             & 0                     \\
stars         &  6             & 2                     \\
QSOs          &  3             & 1                     \\
no $z$        &  20            & 5                     \\
\end{tabular}
\caption{Comparison for different redshift ranges between objects
  falling within Steidel et al.'s selection box and objects selected
  in Steidel et al.'s box which also lie within the CFDF selection box
  (Equation~\ref{eq:selbox}) for $I_{AB}<24.5$.}
\label{tab:cfdfstei}
\end{center}
\end{table}

In total we retrieve $59.6\%$ (31/52) of galaxies at $2.9<z<3.5$ after
applying our selection box. Given that the redshift distribution of
the two samples is different (with mean redshifts of
$\overline{z}=3.04$ and $\overline{z}=3.2$ respectively) this is to be
expected, assuming the underlying distributions are Gaussian with the
same dispersion.
  
Although the photometric selection of the Steidel et al. sample is
different from ours, we can attempt to estimate the amount of
contamination in our catalogue by galaxies outside our redshift range
after the application of our selection box. Inside our selection box,
galaxies at lower redshifts ($2.0<z<2.9$) amount to $25\%$ of the
total.  Spectroscopically identified stars account for a further
$3.8\%$ of objects, in broad agreement with the results of our
simulated catalogues. However, we note that the spectroscopic sample
contains no objects with $z<2$, in disagreement with our simulations.
Finally, $9.6\%$ of our candidates have no redshift.
  
Furthermore, the full spectroscopic catalogue, there are no objects
with $z<2$; however, in $\sim18.5\%$ candidates have no measured
redshift.  Determining redshifts for galaxies in the range $1<z<2$ is
difficult, so it is possible some of these unidentified objects {\it
  could} be galaxies in this redshift range. But as the main fraction
of these object simply have not been attempted yet, this couldn't
account for some of the $\sim 17\%$ of contamination by interlopers
with redshift $z<2$ indicated by our simulations at $I_{AB}<24.5$.

Although our $U$ data is approximately as deep as Steidel et al.'s,
our $BI$ data is somewhat shallower than their $G \mathcal{R}$ images.
For example, detection limits of the Steidel et al. data are
approximately $(U_nGR)_{AB}\sim 27.3,27.3,26.8$
\citep{2003ApJ...584...45A} compared CFDF limits of
$(UBVI)_{AB}=27.0,26.4,26.4,25.6$ ($3\sigma$ limits, $3''$ diameter
aperture, 03hr field; see paper I for more details).  At fainter
magnitudes the shallowness of our $B$ images is expected to increase
our contamination by lower-redshift galaxies. This can explain the
discrepancy between our raw number counts and the number counts of
Steidel et al. as shown in Figure~\ref{fig:ngalcounts}, and the fact
that they are in good agreement after correction from contamination in
our sample.

In summary, we estimate that our sample is contaminated at the level
of $15\%$ to $30\%$ between $I_{AB}<23.5$ and $I_{AB}<24.5$
respectively.  Our selection box allows us to recover a large fraction
of simulated Lyman-break galaxies, ranging from $95\%$ to $80\%$
between $I_{AB}<23.5$ and $I_{AB}<24.5$ respectively.  Comparisons
with a large sample of galaxies with spectroscopic redshifts
(preselected, however, using a different photometric criterion from
ours) indicate we recover, in this case, $\sim 60\%$ of the
Lyman-break galaxies. We attribute this discrepancy to the different
underlying redshift distributions for the two photometrically selected
samples.

\section{Clustering of the Lyman-break galaxies}
\label{sec:clust_LBG}

\subsection{Estimating $\omega(\theta)$ and $A_\omega$}
\label{sec:estim-LBG}

To measure $\omega(\theta)$, the two-point projected galaxy correlation
function, we use the \citet{LS} estimator,
\begin{equation}
\omega ( \theta) ={\mbox{DD} - 2\mbox{DR} + \mbox{RR}\over \mbox{RR}},
\label{eq:ls}
\end{equation}
where the DD, DR and RR terms refer to the number of data-data,
data-random and random-random galaxy pairs having angular separations
between $\theta$ and $\theta+\delta\theta$.

In the weak clustering regime this estimator has a nearly Poissonian
variance \citep{LS}, 
\begin{equation}
\delta\omega ( \theta) = \sqrt{\frac{1+\omega(\theta)}{\mbox{DD}}}.
\label{eq:err-ls}
\end{equation}
Section~\ref{sec:rob-meas} addresses the reliability of this error
measurement for our present dataset.  To determine $A_{\omega}$, the
amplitude of $\omega(\theta)$ at 1 degree, we assume that
$\omega(\theta)$ is well represented by a power-law of slope $\delta$,
i.e. $\omega ( \theta) = A_{\omega}\theta^{-\delta}$
\citep{1977ApJ...217..385G}.  In what follows, we assume $\delta=0.8$;
in Section~\ref{sec:measuring-slope} we explore this assumption in
more detail. This fitted amplitude must be adjusted to take into
account the ``integral constraint'' correction, which arises from the
fact that the mean background density of galaxies is estimated from
the sample itself. We estimate this term as follows \citep{RSMF},
\begin{equation}
C = {1 \over {\Omega^2}} \int \! \int \omega(\theta) d\Omega_1 d\Omega_2,
\label{eq:C}
\end{equation}
where $\Omega$ is the area subtended by the survey field. To determine
$C$ we numerically integrate this expression over each field,
excluding masked regions. We find $C\simeq4.2A_{\omega}$ for the 14hr
and 03hr fields. For the 22hr field, which has half the coverage, we
derive $C\simeq5.5A_{\omega}$.  Then we determine $A_\omega$ fitting
the expression:
\begin{equation}
\omega_{obs} ( \theta) = A_{\omega}\theta^{-\delta} - C.
\label{eq:wobsfit}
\end{equation}

\begin{figure}
  \resizebox{\hsize}{!}{\includegraphics{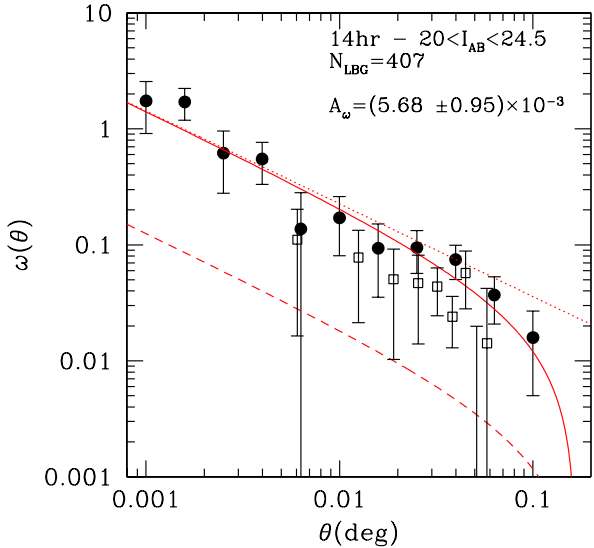}}
\caption{The amplitude of the angular correlation function, $\omega(\theta)$,
  as a function of the angular separation in degrees, for a
  $20<I_{AB}<24.5$ limited Lyman-break sample of the CFDF-14hr field
  (filled black circles).  Errorbars represent normal Poisson errors
  (Equation~\ref{eq:err-ls}).  The solid line shows the fitted
  correlation amplitude, derived using Equation~\ref{eq:wobsfit} and
  assuming a power-law slope of $\delta=0.8$ and a value of
  $C=4.2A_{\omega}$ for the integral constraint term. The dotted line
  shows the fitted power-law without the integral constraint
  correction. The long dashed line shows the fitted correlation
  amplitude (from paper I) for field galaxies selected with the same
  limiting magnitude. Open squares are the $\omega(\theta)$
  measurements from the \citet{1998ApJ...503..543G} $\mathcal{R}<25.5$
  selected Lyman-break sample.}
\label{fig:omtheta}
\end{figure}

Figure~\ref{fig:omtheta} shows the angular correlation function,
$\omega(\theta)$, as a function of the angular separation in degrees
for Lyman-break galaxies with $20.0<I_{AB}<24.5$ in the CFDF-14hr
field. Here the errorbars have been estimated with the normal Poisson
errors (Equation~\ref{eq:err-ls}). The fitted amplitude derived from
Equation~\ref{eq:wobsfit} is represented by the solid line. The long
dashed line shows the $A_\omega$ value computed from the CFDF field
galaxy sample at the same limiting magnitude (from Paper I). The
$A_\omega$ for the Lyman-break sample is $\sim 10$ higher than the
field galaxy sample. In Table~\ref{tab:resultr0} we summarise our
Lyman-break $A_{\omega}$ measurements for a range of limiting
magnitudes in the CFDF.

In Figure~\ref{fig:omtheta} we compare our measurements of
$\omega(\theta)$ to those of \cite{1998ApJ...503..543G}. As the
largest CFDF fields are $\sim 9$ times larger than those used in this
study, our measurements of $\omega(\theta)$ cover a much larger range
of angular separations. We note that our amplitude measurements are
$\sim 2$ times larger those of \citeauthor{1998ApJ...503..543G}; we
expect this arises from the greater depth of the
\citeauthor{1998ApJ...503..543G} study compared to the CFDF. To test
that the origin of this discrepancy in amplitude did \textit{not}
arise from inhomogeneities within our fields, we extracted, from each
CFDF field, sub-fields covering the same $9'\times 9'$ area as
subtended by the \citeauthor{1998ApJ...503..543G} work. In total we
extracted 21 fields of these dimensions. We fitted each sub-field
individually and found a median correlation amplitude over all fields
of $(6.9\pm5.1)\times10^{-3}$, which is in good agreement with the
full field value of $(7.4\pm1.0)\times10^{-3}$ quoted in
Table~\ref{tab:resultr0}. The results of this test are consistent
with the simulations carried out in paper I in which we demonstrated
that measurements of $\omega(\theta)$ for $I_{AB}-$ limited samples in
the CFDF are unaffected by sensitivity variations across the mosaics
to at least $I_{AB}\sim25$. Finally, we also note that our
measurements of $\omega(\theta)$ follow a power-law behaviour over the
entire range $0.001^{\circ} < \theta < 0.02^{\circ}$ accessible to our
survey and there is no evidence of an excess of power on large scales
(with the exception of the 03hr field), as one might expect if
residual inhomogeneities existed within individual field.

\begin{figure}
  \resizebox{\hsize}{!}{\includegraphics{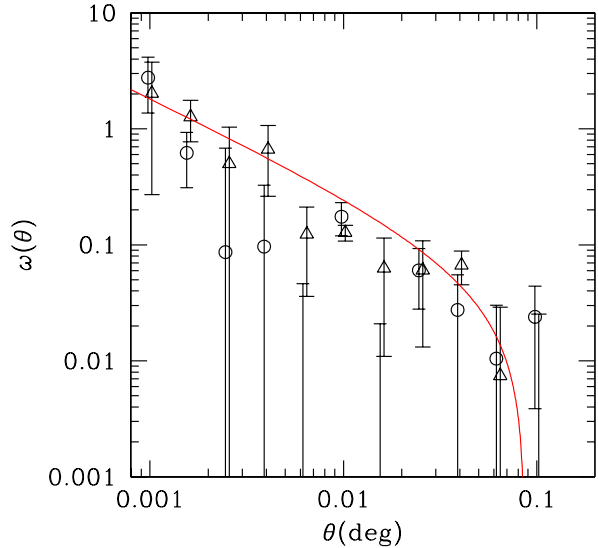}}
\caption{The quadrant-averaged Lyman-break correlation function
  $\omega(\theta)$ for galaxies in the 03hr and 22hr fields (triangles
  and circles respectively). The amplitude of these error bars
  corresponds to the quadrant-to-quadrant variation in
  $\omega(\theta)$.  The solid line shows the fitted correlation
  amplitude for Lyman break galaxies in the 14hr field with a fixed
  slope applied for one quadrant (Figure~\ref{fig:omtheta}).}
\label{fig:byquadrant}
\end{figure}

We also measured $\omega(\theta)$ in four separate sub-areas on each
of our three fields. Each sub-areas corresponds to the size of the
individual $U$-band pointings. In each sub-area we measure
$\omega(\theta)$ separately and then determine the mean and the
variance: this is illustrated in Figure~\ref{fig:byquadrant}.
Measuring $\omega(\theta)$ is these sub-areas is more challenging as
the numbers of galaxies involved is much smaller. However, the fitted
amplitudes in each of these sub-areas agrees very well with the
full-field values presented in Table~\ref{tab:resultr0}.

\subsection{Measuring the slope}
\label{sec:measuring-slope}

Is the slope of the $\omega(\theta)$ for Lyman-break galaxies really
$\delta=0.8$? In earlier works
\citep{1998ApJ...505...18A,1999MNRAS.310..540A}, a value of
$\delta=0.8$ was used based on results from local large surveys
\citep{1977ApJ...217..385G}. In contrast, \citet{1998ApJ...503..543G}
measured $\delta=1.0\pm0.3$. The large angular coverage of the CFDF
fields allow estimate $\delta$; in Figure~\ref{fig:omtheta}, we can
easily detect power in $\omega(\theta)$ to scales of $\sim
0.1^{\circ}$, making it possible to place constraints on the joint
values of $A_\omega$ and $\delta$.

\begin{figure}
\resizebox{\hsize}{!}{\includegraphics{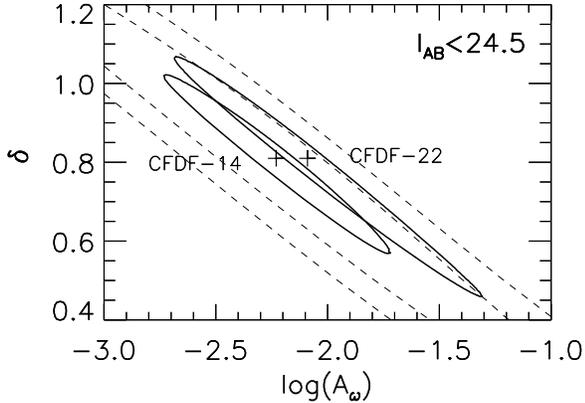}}
\caption{ Contours of $\chi^2$ for the mean $\omega(\theta)$ computed for
  Lyman-break galaxies selected with $I_{AB}<24.5$ in the CFDF-14hr
  and CFDF-22hr fields.  The plus symbols show the best-fitting
  amplitudes and slopes, the contours correspond to the 1$\sigma$
  (thick contours) and 3$\sigma$ confidence levels.}
\label{fig:Aw-delta}
\end{figure}

To estimate the best-fitting values of $A_\omega$ and $\delta$ we
carry out a $\chi^2$ minimisation on the values of $\omega(\theta)$
determined for all fields, similar to that described in Paper I.  This
computation accounts for the dependence of the integral constraint $C$
with the slope $\delta$. Figure~\ref{fig:Aw-delta} shows the fit for
two of our three fields; our data provides an approximate constraint
on $\delta$. We find the mean of the best fitted slopes is
$\delta=0.81^{+0.21}_{-0.24}$ for the CFDF-14hr field and
$\delta=0.81^{+0.25}_{-0.35}$ for the CFDF-22hr field for
$I_{AB}<24.5$ (we were not able to fit simultaneously both for the
slope and amplitude on the CFDF-03hr field).

To summarise, our clustering measurements are broadly consistent with a
power-law of slope $\delta \sim 0.8$ over all the magnitude ranges
accessible to our survey.  Our data do not provide any strong evidence
for slopes shallower or steeper than this value, as suggested by other
authors \citep{2001ApJ...550..177G,2003ApJ...584...45A}.

\subsection{The comoving correlation length $r_0$}
\label{sec:meas-r0}

We use the spatial correlation function \citep{1977ApJ...217..385G}, to
derive $r_0$, the comoving galaxy correlation length  based on our
angular clustering measurements, given by
\begin{equation}
\xi(r,z) = \left( { \frac{r}{r_0(z)} } \right)^{-\gamma},
\label{eq:xi}
\end{equation}
where $\gamma=1+\delta$. The redshift dependence is included in the
comoving correlation length $r_0(z)$.

Using the relativistic Limber equation \citep{P80,1999MNRAS.306..988M},
we can derive the correlation length $r_0$ from the correlation
amplitude $A_{\omega}$, providing we can estimate a redshift
distribution for our sources.  In what follows we assume that our
Lyman-break redshift distribution is well described by a top-hat
function spanning the interval $2.9<z<3.5$; however, our results are
unchanged if we use a Gaussian redshift distribution covering the same
interval. 

Could our adopted redshift distribution be modified by the presence of
interlopers?  Assuming a Gaussian distribution of Lyman-break galaxies
centred on $2.9<z<3.5$ with $\overline{z}=3.2$ and $\sigma_z=0.3$, we
estimate in the following manner the effect that $30\%$ of
contamination on $\overline{z}$: first, we assume the redshift
distribution of the interlopers is also a Gaussian with
$\overline{z}=2.5$ and $\sigma_z=0.3$.  Next, adding $30\%$ of these
object to our reference distribution we find $\overline{z}=3.0$ with
$\sigma_z=0.4$, i.e. a variation of $5\%$.  Interlopers at lower
redshifts, $\overline{z}=1.8$ and $\sigma_z=0.3$ produce a $10\%$
variation in $\overline{z}$.  These numbers are unchanged if instead we
assume top-hat interloper distribution. Based on this discussion we
adopt a $10\%$ as upper limit of to our uncertainty in $\overline{z}$.

In Table~\ref{tab:resultr0} we present the values of the comoving
correlation length $r_0$ of Lyman-break galaxies with
$20.0<I_{AB}<24.5$. If we incorporate the uncertainty in $\overline z$
outlined above, an extra error of $\pm 0.2$ for the samples with
$20.0<I_{AB}<24.5$, and of $\pm 0.4$ for $20.0<I_{AB}<23.5$ must be
added.  Results are shown for three cosmologies: Einstein-DeSitter
($\Omega_0=1.0, \Omega_{\Lambda}=0.0$), open ($\Omega_0=0.2,
\Omega_{\Lambda}=0.0$) and $\Lambda$-flat ($\Omega_0=0.3,
\Omega_{\Lambda}=0.7$). We present correlation lengths computed for
each field and for the mean of the three fields. We also show the
results for two-parameter fits (slope and amplitude) for the 14hr and
22hr fields, and also for a fixed slope $\delta=0.8$ for the 03hr
field and for the mean of all fields.  Errors were computed using the
Poissonian statistics (Equation~\ref{eq:err-ls}).

We note that our two-parameter fits for $r_0$ and slope are not
consistent with those of \citet{2003ApJ...584...45A}
($r_0=(4.0\pm0.3)h^{-1}$ Mpc and $\delta=0.55\pm0.15$); these
measurements fall outside the error ellipses plotted in
Figure~\ref{fig:Aw-delta}.  Two phenomena could explain this
discrepancy: firstly the sample of \citeauthor{2003ApJ...584...45A} is
slightly fainter than ours (which could produce a shift of the contour
in Figure~\ref{fig:Aw-delta} to the right -- see
Section~\ref{sec:meas-seg}) and secondly their sample is a
spectroscopic one and is expected to have a lower level of
contamination than ours.

To summarise, for a $\Lambda$-flat cosmology, and for $\delta=0.8$, we
derive for the full $20.0<I_{AB}<24.5$ sample $r_0=(5.9\pm0.5)h^{-1}$
Mpc, averaged over all three fields.  For the two-parameter fits, we
find $r_0 = (5.3^{+6.8}_{-2.2})h^{-1}$~Mpc with $\delta =
0.81^{+0.21}_{-0.24} $ and $r_0 = (6.3^{+17.9}_{-2.8})h^{-1}$~Mpc with
$\delta = 0.81^{+0.25}_{-0.35}$ respectively.

\subsection{Possible dependence on apparent magnitude of the correlation amplitude and length}
\label{sec:meas-seg}

\begin{figure}
  \resizebox{\hsize}{!}{\includegraphics{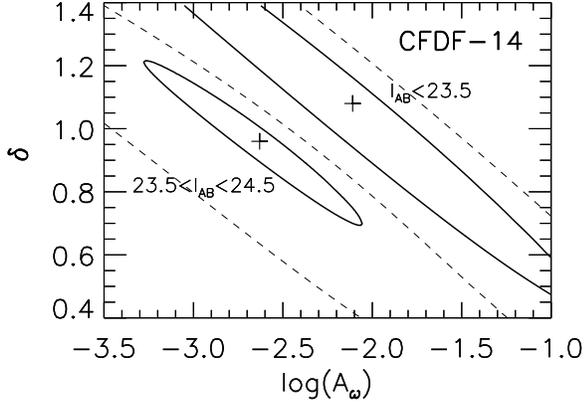}}
\caption{Contours of $\chi^2$ for $\omega(\theta)$ for two
  subsamples of Lyman-break galaxies with $20.0<I_{AB}<23.5$ and
  $23.5<I_{AB}<24.5$ in the CFDF-14hr field.  The plus symbol shows
  the best-fitting amplitudes and slope, and two contours correspond
  to the 1$\sigma$ (thick contours) and 3$\sigma$ confidence levels.}
\label{fig:Aw-delta-seg}
\end{figure}

In the 14hr field, our $r_0$ measurements indicate samples with
fainter limits magnitudes have \textit{lower} values of $r_0$.
Comparing our brightest $20.0<I_{AB}<23.5$ and our faintest
$23.5<I_{AB}<24.5$ samples, we detect this effect with a $3 \sigma$
confidence, as shown in the Figure~\ref{fig:Aw-delta-seg}.  For these
two sub-samples we also find approximately the same values of the
slope. In Table~\ref{tab:resultr0}, we show the values of the
correlation amplitude and length for the two magnitude-limited
samples, with the slope fixed to $\delta=0.8$ and not fixed.

We note that this dependence of clustering strength with luminosity is
also observed at lower redshifts
\citep{2002MNRAS.332..827N,Bud03_astroph}. Moreover, recent results
from the SDSS \citep{Bud03_astroph} demonstrate that the slope of
galaxy correlation function is independent of galaxy absolute
luminosity, consistent with our observations.

\renewcommand{\arraystretch}{1.2}
\begin{table*}[htbp]
\begin{center}
\begin{tabular}{*{7}{c}}
{\bf Field} & magnitude limit &  $A_{\omega}$(1 deg) & $\delta$ & $r_0$ 
            ($h^{-1}$ Mpc) & $r_0$ ($h^{-1}$ Mpc) & $r_0$ ($h^{-1}$ Mpc) \\
 & ($I_{AB}$) & $\times10^{-3}$ & & $\Omega_{0}=1.0$, & $\Omega_{0}=0.2$, & $\Omega_{0}=0.3$, \\
 &           &                 & & $\Omega_{\Lambda}=0.0$ & $\Omega_{\Lambda}=0.0$ &
                                                                $\Omega_{\Lambda}=0.7$ \\
\hline               
 &  &  &  &  &  & \\
{\bf CFDF-14} & 20.0--24.5 &  5.9$^{+13.2}_{-4.0}$ & {\bf
            0.81$^{+0.21}_{-0.24}$} & 3.2$^{+4.0}_{-1.2}$ &  3.6$^{+4.4}_{-1.3}$ & 5.3$^{+6.6}_{-2.0}$ \\
          & 20.0--23.5 & 7.8$^{+166.0}_{-7.7}$ & 
            1.08$^{+0.84}_{-0.66}$ & 5.8$^{+59.8}_{-2.8}$ &  6.8$^{+69.3}_{-3.2}$ & 9.5$^{+97.7}_{-4.5}$ \\
          & 23.5--24.5 & 2.3$^{+6.4}_{-1.7}$ &
            0.96$^{+0.25}_{-0.26}$ & 2.6$^{+3.2}_{-1.0}$ &  3.0$^{+4.2}_{-1.1}$ &  4.3$^{+6.1}_{-1.6}$ \\
{\bf CFDF-22} & 20.0--24.5 &  8.1$^{+40.9}_{-6.0}$ & {\bf
            0.81$^{+0.25}_{-0.35}$} & 3.8$^{+10.7}_{-1.6}$ &
            4.2$^{+11.8}_{-1.7}$ & 6.3$^{+17.7}_{-2.6}$ \\
{\bf CFDF-03} & 20.0--24.5 &  8.6$\pm$0.6 & 0.8 & 3.9$\pm$0.2 & 4.3$\pm$0.2 &  6.4$\pm$0.3 \\
 &  &  &  &  &  & \\  
{\bf CFDF mean} & 20.0--24.5 &  7.4$\pm$1.0 & 0.8 & 3.6$\pm$0.3 &
            3.9$\pm$0.3 &  {\bf 5.9$\pm$0.5} \\
{\bf CFDF-14} & 20.0--23.5 & 24.9$\pm$7.9 & 0.8 & 7.0$\pm$1.2 &
            7.7$\pm$1.4 & {\bf 11.6$\pm$2.0} \\
          & 23.5--24.5 & 5.4$\pm$1.1 & 0.8 & 3.0$\pm$0.3 & 3.3$\pm$0.4
            & {\bf 5.0$\pm$0.6} \\
 &  &  &  &  &  & \\  
\end{tabular}
\caption{The amplitude of $\omega(\theta)$ at 1 degree, $A_{\omega}$,
  the slope $\delta$ and the comoving correlation length $r_0$
  (in~$h^{-1}$~Mpc), for each field and for different magnitude
  limited samples considered in this paper. $r_0$ measurements are
  computed for three standard cosmological models.  To derive $r_0$ we
  assume a top-hat redshift distribution centred at $\overline z =
  3.2$ and the best fitted value of the slope. Result marked as CFDF
  mean are computed from the mean over all three fields.  The error
  bars shown correspond to Poisson error bars. An extra systematic
  errorbar arising from our uncertainty in the underlying redshift
  range of our Lyman-break sources of $\pm 0.2$ for the entire faint
  samples and of $\pm 0.4$ for the bright sample should be added. (Our
  principal results are highlighted in bold.)}
\label{tab:resultr0}
\end{center}
\end{table*}
\renewcommand{\arraystretch}{1}

\subsection{Effect of contaminants on $A_\omega$ and $r_0$}
\label{sec:effect-interl-a_om}

To estimate the effect the contaminating population has on our
measurements of $r_0$ and $A_\omega$, we must make some assumptions of
their clustering properties. In the case of the stellar contaminants,
this is easy; however, for the interloper population it is less clear.
Our selection criterion of $(V-I)<1$ eliminates $z\sim1.5$ bright
ellipticals which might produce spuriously high correlations
(additionally, we find no trend in our samples of $(V-I)$ with
$I_{AB}$ magnitude). Moreover, our simulations indicate that most of
the interloper population lies at $z\sim2$.

Assuming \textit{all} the contaminants are unclustered, we can derive
upper limits of the effect on the clustering. We find a fraction of
contamination (by lower-redshift interlopers and stars) of
$f\simeq0.15$ for $20<I_{AB}<23.5$ and $f\simeq0.3$ for the fainter
$20<I_{AB}<24.5$ (Section~\ref{sec:robustn-select-box}). If these
objects are not clustered, the estimates of clustering amplitudes
$A_{\omega}$, assuming a fixed slope of $\delta=0.8$, have to be
readjusted by a factor $1/(1-f)^2\simeq1.38$ for the brighter sample
and $1/(1-f)^2\simeq2.04$ for the fainter one.  This implies factors
of $\simeq1.20$ and $\simeq1.49$ respectively for the correlation
lengths $r_0$ for bright and faint samples (in our fainter magnitude
bins, the interloper population is composed primarily of galaxies,
which are more strongly clustered than stars but less strongly
clustered than the Lyman-break population; this may further reduce the
factor of 1.49).

An empirical way to estimate the effect of the contamination on our
measurements is to replace a fraction of our candidates by objects
extracted from the whole catalogue. We carried out this exercise for
the 14hr field by replacing 30\% of the objects with $20<I_{AB}<24.5$
and 15\% for $20<I_{AB}<23.5$, computing clustering with a slope of
$\delta=0.8$ and for a $\Lambda$-flat cosmology. In the first case we
find $r_0=(10.3\pm2.2)h^{-1}$~Mpc, i.e.  a factor of $\simeq 1.13$
times lower, and in the second case $r_0=(4.1\pm0.5)h^{-1}$~Mpc, i.e.
a factor of $\simeq 1.22$ times lower. Of course in this experiment we
cannot control the nature of the contaminants but these results
indicate that this level of contamination could not produce the
$3\sigma$ segregation effect reported in section~\ref{sec:meas-seg}.

\subsection{Are Poisson errors appropriate for the Landy and Szalay
  estimator?}
\label{sec:rob-meas}

\begin{figure}
  \resizebox{\hsize}{!}{\includegraphics{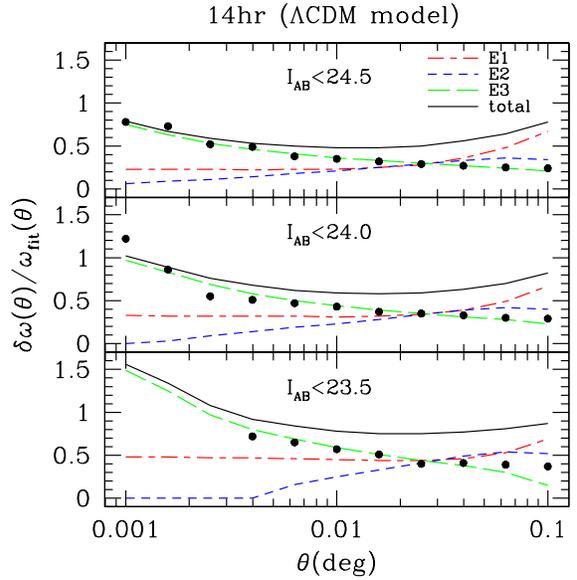}}
\caption{Comparison of the different errors contributing to the global cosmic
  errors as a function of angular scales for three different magnitude
  cuts of the CFDF-14hr sample.  The main source of errors are: the
  finite volume error $E_1^{1/2}$ (long-short-dashed line); the
  discreteness errors $E_2^{1/2}$ (short-dashed line), $E_3^{1/2}$
  (long-dashed line), and the total cosmic error
  $E^{1/2}\equiv(E_1+E_2+E_3)^{1/2}$ (solid line) (see
  \citealp{1994ApJ...424..569B}). Filled black circles are the
  Poissonian errors estimated from Equation~\ref{eq:err-ls}. The
  analytical errors are computed using a ``$\Lambda$CDM bias model''
  (see Section~\ref{sec:rob-meas}).}
\label{fig:errAw}
\end{figure}

In this section we investigate if the errors in the Landy and Szalay
estimator (Equation~\ref{eq:err-ls}) can be reliably described by
Poissonian statistics. In doing this, we neglect other contributions,
such as the finite size of the survey and the clustered nature of the
galaxy distribution. We estimate here the relative amount of the
various contributions to the error budget, using the analytical
expression derived by \citet{1994ApJ...424..569B}.  This expression
has three terms: one reflecting the finite volume error ($E_1$: ``{\it
  cosmic variance}''), which is independent of the number of galaxies,
and two others related to the discrete nature of the galaxy catalogue:
the first one appears only in the case of correlated sets of points
($E_2$, which cancels if $\omega \rightarrow 0$) and the second one
includes the pure Poisson error ($E_3$).  The calculation of the
cosmic error requires prior knowledge of higher-order statistics
($S_3$ and $S_4$) as well as their redshift behaviours. We follow the
recipes described in \citet{2000chr..conf..153C} and
\citet{2002MNRAS.329..355A} to perform this computation. Of course our
total error budget will be dominated by the effects of systematic
errors arising from our imperfect knowledge of the source redshift
distribution and the precise quantity of contaminants in our sample,
as we have discussed extensively elsewhere in our paper.

In Figure~\ref{fig:errAw} we show the relative magnitudes of the three
components $E_1^{1/2}$ (short-long dashed lines), $E_2^{1/2}$ (short
dashed lines) and $E_3^{1/2}$ (long-dashed lines) and the total error
($E^{1/2}= \sqrt{E_1+E_2+E_3}$, solid lines) as a function of the
angular scale, $\theta$, for three limiting magnitudes.  The results
are shown for the ``$\Lambda$CDM bias model'' described in
\citet{2002MNRAS.329..355A}.  The bias values used in this analysis
for the different samples are given in Table 5.  The theoretical
estimates are compared to the observed errors of the CFDF-14hr field
($\delta\omega/\omega_{fit}$).

The behaviour of the observed errors matches closely the Poisson term
$E_3$ at all angular scales for each of the three magnitude limited
samples.  At $\theta \le 0.02^{\circ}$, $E_3$ dominates the total
error. The contribution of the finite volume error $E_1$ starts to
play a significant role at relatively large scales: $0.04^{\circ} \le
\theta \le 0.1^{\circ}$. The contribution of $E_2$ is never dominant
at any scale.  For $0.001^{\circ}<\theta<0.02^{\circ}$, our analysis
shows that the total cosmic error ($E$) is dominated by Poisson noise
($E_3$) and the amplitude of $E(\theta\sim0.02^{\circ})^{1/2}$ is not
more than a factor 1.6 larger than the amplitude of $E_3^{1/2}$.  This
result justifies the choice of using the nearly Poissonian errors.

\section{Discussion and comparison with theory}
\label{sec:comp-with-theor}

\subsection{Introduction}
\label{sec:disc-int}

In this section we compare our measurements of the galaxy correlation
length $r_0$ (in $h^{-1}$~Mpc) with those of other authors and we
interpret these derived values in terms of several simple models.
Throughout this section, unless stated otherwise, all measurements of
$r_0$ are presented for a $\Lambda$-flat cosmology ($\Omega_0=0.3$ and
$\Omega_{\Lambda}=0.7$). When necessary, we transform measurements from
other authors to this cosmology using the equations presented in
\cite{2000MNRAS.314..546M}. As we have already demonstrated in
section~\ref{sec:measuring-slope}, our measured slopes are consistent
with $\delta = 0.8$; to comparing our results with literature
measurements and models, we use this corresponding value of the slope.

\subsection{Tracing the evolution of $r_0$ with redshift}
\label{sec:spatial-modz}

\begin{figure*}[htb]
  \begin{center}
    \resizebox{\smallysize}{!}{\includegraphics{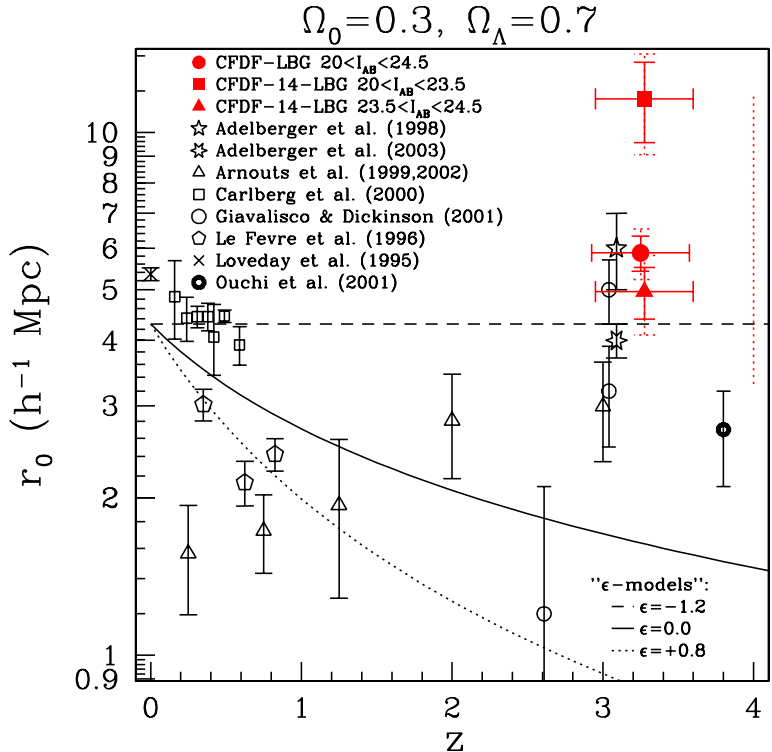}}
  \end{center}
\caption{The comoving correlation length, $r_0$ (in $h^{-1}~$Mpc), for
  three samples of Lyman-break galaxies in the CFDF with the slope
  value fixed to $\delta=0.8$ (symbols slightly offset for clarity).
  The circle, square and the triangles represent the mean correlation
  length for each of the three fields, the correlation length for a
  magnitude-limited sample of the CFDF-14hr fields with
  $20.0<I_{AB}<23.5$ and the correlation length for a
  magnitude-limited sample of the CFDF-14hr fields with
  $23.5<I_{AB}<24.5$ respectively. We plot a range of $r_0$
  measurements from the literature, which are described in detail in
  Section~\ref{sec:spatial-modz}. The horizontal error bars represent
  an uncertainty of $10\%$ in the mean redshift. The solid vertical
  errorbars on $r_0$ are computed using Poisson statistics.  Dotted
  vertical error bars represent the addition of this redshift error to
  the Poissonian component. The dotted line on the right part of the
  figure represents the errorbars for the CFDF-14hr whole sample when
  we fit for both the slope and amplitude.}
\label{fig:r0z_l}
\end{figure*}

In Figure~\ref{fig:r0z_l}, we plot the comoving correlation length
$r_0$ of Lyman-break galaxies in the CFDF. The filled circle shows the
mean measurement for all fields, for $20.0<I_{AB}<24.5$; the filled
square and filled triangle shows measurements at $20.0<I_{AB}<23.5$
and $23.5<I_{AB}<24.5$ respectively for the CFDF-14hr field.  These
measurements are shown at the mean assumed redshift of the CFDF
Lyman-break sample $\overline{z}=3.2$. For clarity each of the samples
is slightly offset from each other. In addition, the dotted line on
the right of this Figure represents the errorbar for the CFDF-14hr
$20.0<I_{AB}<24.5$ sample when both $\delta$ and the amplitude are
fitted, and corresponds to a projection of the $1\sigma$ contour plot
shown in Figure~\ref{fig:Aw-delta} along the amplitude axis.

For comparison we also show $r_0$ measurements for the local Universe
from the Stromlo-APM survey \citep{1995ApJ...442..457L}. Clustering
measurements from the $I_{AB}<22.5$~selected CFRS and the CNOC
absolute magnitude-limited~$M_R^{k,e}<-20$ surveys provide
measurements of the evolution of clustering to $z<1$
\citep{1996ApJ...461..534L,2000ApJ...542...57C}.  We also show an
average of measurements based on photometric redshift studies of the
HDF-N and -S \citep{1999MNRAS.310..540A,2002MNRAS.329..355A}; galaxies
in this study have $I_{AB}\leq28$ (clustering measurements at $z\sim4$
from this study are not shown because of the very small numbers of
galaxies in this redshift bin). Finally, correlation length derived
for $z\sim3.8$ galaxies with $i'_{AB}\leq26$ in the Subaru deep field
\citep{2001ApJ...558L..83O} is shown.

A comparison of previous clustering measurements of Lyman-break
galaxies are also presented. We note that in the literature there are
several different analyses of the same dataset or supersets of the
same dataset (either the HDF fields or the fields analysed by Steidel
and collaborators). The open stars show measurements from
\citet{2003ApJ...584...45A}, who fit for both slope and amplitude; the
\cite{1998ApJ...505...18A} measurement was carried out on a subsample
of this, with the slope fixed to $\delta=0.8$ (for clarity those
measurements were slightly offset).  The three open circles show the
clustering measurements from \cite{2001ApJ...550..177G}; the upper
circle represents their $r_0$ measurement from a
$\mathcal{R}_{AB}\lesssim25.1$ spectroscopically selected sample of
Lyman-break galaxies (their ``SPEC'' sample), another subset of the
\citet{2003ApJ...584...45A} sample. The middle open circle is the
\citeauthor{2001ApJ...550..177G} $\mathcal{R}_{AB}<25.5$
photometrically selected Lyman-break sample (the ``PHOT'' sample), and
the lower circle is \citeauthor{2001ApJ...550..177G}'s measurement of
Lyman-break galaxies with $V_{AB\,606}<27$ in the HDF. We caution that
the \citeauthor{2001ApJ...550..177G} use a slope of $\delta\sim1.2$,
different from this work.  This explains the discrepancy between the
HDF clustering measurement by \citeauthor{2001ApJ...550..177G} and
that of average HDF-N and -S values from
\citeauthor{1999MNRAS.310..540A}, who computed fitted correlation
quantities assuming $\delta=0.8$.

Figure~\ref{fig:r0z_l} also shows ``$\epsilon$-model'' predictions,
i.e, $r_0(z) = r_0(z=0) (1+z)^{-(3+\epsilon-\gamma)/\gamma}$, for
different values of $\epsilon$, scaled arbitrarily to the value of
$r_0=4.3h^{-1}$ Mpc at redshift $z=0$ \citep{1977ApJ...217..385G}. In
this simple prescription, three values of $\epsilon$ are normally
considered: $\epsilon=-1.2$ for a slope $\gamma=1.8$, corresponding to
clustering fixed in comoving coordinates; $\epsilon=0$, representing
clustering fixed in proper coordinates; and $\epsilon=0.8$ which
corresponds to the predictions of linear theory, for an
Einstein-DeSitter cosmology.

Taken together, these measurements present no clear picture of how
$r_0$ evolves with redshift; for the CNOC survey, it appears that
clustering is approximately fixed in comoving coordinates up to
$z\sim0.6$, whereas the results of the CFRS study indicate $r_0$
declines to $z\sim1$. The HDF $r_0$ measurements appear to increase
gradually over the entire redshift interval shown in our graph, and
are always below the high-redshift values estimated from the CFDF. A
number of separate factors contribute to this disparity: firstly, as
we have highlighted, each individual sample has a different selection
criterion; for example, galaxies at $z<1$ from the HDF samples are
much fainter and less numerous than those selected in the CFRS survey.
It is clear from local spectroscopic surveys that galaxy clustering is
a sensitive function of spectral type and intrinsic luminosity
\citep{1995ApJ...442..457L,1999MNRAS.310..281L,2002MNRAS.332..827N}.
Secondly, the field of view and the comoving scales probed are very
different between each survey. At $z\sim1$, for example, the HDF
probes only $1h^{-1}$~Mpc, and this comoving scale increases at higher
redshifts. Lastly, all surveys are subject to sampling and cosmic
variance effects.

Precisely because of the effects outlined above it is difficult to
directly compare our measurements of $r_0$ for Lyman-break galaxies to
those of other authors. As mentioned previously, an additional
uncertainty is that not all authors adopt the same value of the slope
$\delta$, although the strong covariance between $A_\omega$ and
$\delta$ allows us to estimate approximately the effect a changing
slope will have on the fitted amplitude (see
Figure~\ref{fig:Aw-delta}). Furthermore, all previous measurements of
clustering at high redshift, based on photometric samples such as
ours, are for fainter magnitudes than the faintest CFDF sample. Given
the observed segregation of clustering amplitude with apparent
magnitude observed in the CFDF-14hr field, we would expect these
previous studies to measure a lower amplitude than our work, and this
is indeed what is observed.  The photometric sample of
\citet{2001ApJ...550..177G}, reaching a half-magnitude fainter than
our faintest sample, displays a clustering amplitude approximately
twice as low as our faintest bin.  However, the {\it spectroscopic}
Lyman-break samples of \citet{1998ApJ...505...18A} and
\citet{2001ApJ...550..177G} have approximately the same magnitude
limits as our work, and we agree quite well with these measurements.

\subsection{The surface density dependence of $r_0$}
\label{sec:spatial-modn}

\begin{figure}
  \resizebox{\hsize}{!}{\includegraphics{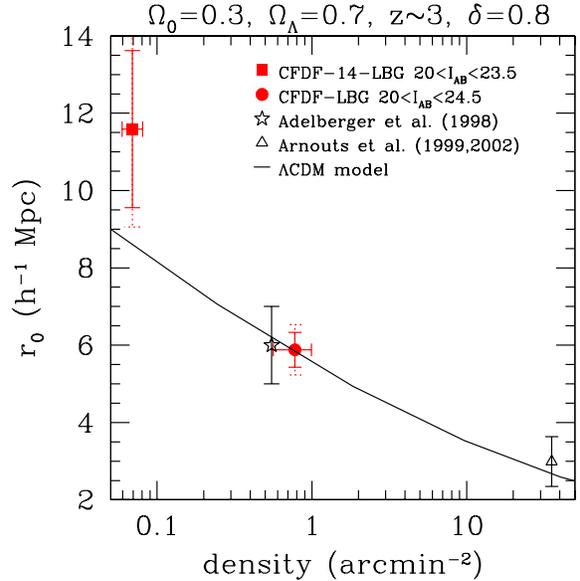}} 
\caption{The comoving correlation length $r_0$ (in $h^{-1}$~Mpc)
  for two magnitude-limited CFDF samples (filled circle symbol for the
  mean over the three fields, and filled square symbol for the
  CFDF-14hr field with $I_{AB}<23.5$), as a function of cumulative
  surface density on the sky.  Measurements from other Lyman-break
  samples (open symbols), and from the mean of HDF-N and -S (open
  triangles) are displayed.  The solid vertical errorbars on $r_0$ are
  computed using Poisson statistics.  Dotted vertical error bars
  represent the addition of the redshift error to the Poissonian
  component.}
\label{fig:r0surf_l}
\end{figure}

In this Section we discuss the dependence of $r_0$ with galaxy surface
density, a relationship which is more amenable to direct modelling,
and discuss in more detail the implications of the segregation of
galaxy clustering with apparent magnitude.

Figure~\ref{fig:r0surf_l} shows the comoving correlation length $r_0$
as a function of surface density for two magnitude limited samples
($20.0<I_{AB}<23.5$ and $20.0<I_{AB}<24.5$) extracted from the
CFDF-14hr and averaged over all three CFDF fields respectively.  Error
bars are computed using Poisson statistics. We added two clustering
measurement of Lyman-break galaxies taken from the literature:
\citet{1998ApJ...505...18A}, and the average of two measurements for
Lyman-break galaxies photometrically selected in the HDF-N and -S
\citep{1999MNRAS.310..540A,2002MNRAS.329..355A} (here we only show
measurements of $r_0$ computed assuming a slope $\gamma=1.8$). At
densities of $\sim1$~arcmin$^{-2}$ our $r_0$ measurements are in
excellent agreement with those of \citeauthor{1998ApJ...505...18A}.
Moreover, our measurements show a trend of increasing correlation
length with decreasing galaxy surface density.

As an attempt to interpret these results, we consider the
$\Lambda$-CDM analytic model of structure evolution presented in
\citet{1999MNRAS.310..540A} and discussed fully in
\citet{1997MNRAS.286..115M} and \citet{1998MNRAS.299...95M} (their
``transient'' model). The relevant cosmological parameters for this
model are given in Table~\ref{tab:resultb} of
\citeauthor{1998MNRAS.299...95M}'s paper. Similar models have also
been presented elsewhere
\citep{1996MNRAS.282..347M,1999MNRAS.304..175M}. In this model, each
Lyman-break galaxy is associated with one dark matter halo.

To briefly summarise the model's main components, we assume that the
clustering of galaxies $\xi_g(r,z)$ is linearly related to the dark
matter clustering $\xi_m(r,z)$ through the linear effective bias
$b^2_{eff}(z)$. The dark matter clustering is computed in the
non-linear regime occupied by our survey using the fitting formulae of
\citet{1996MNRAS.280L..19P} and a power spectrum normalised to
correctly reproduce the present-day abundance of bright clusters. The
effective bias is calculated by integrating the product of the bias
parameter $b(M,z)$ and the \citet{1974ApJ...187..425P} dark matter
halo redshift-mass distribution function over all the masses of halos
larger than a typical minimum mass.  To improve accuracy, the models
use the fitting formulae of \citet{1999MNRAS.308..119S} for these
quantities based on halos identified in a large N-body simulation.
This model is shown as the solid line in Figure~\ref{fig:r0surf_l}.

Despite its simplicity, this model reproduces quite well the observed
strong dependence of $r_0$ on Lyman-break surface density seen in the
CFDF survey, and this agreement continues to very faint $I_{AB}=28$
measurements at surface densities of $\sim 40$~arcmin$^{-2}$ from the
combined HDF measurement.  Previously, such a dependence had not been
unambiguously detected \textit{within} a given survey.

These results argue against models of Lyman-break galaxy clustering
such as the ``bursting'' scenario proposed by
\citet{1999ApJ...523L.109K}, in which Lyman-break galaxies become
visible as a result of stochastic star-formation activity. These
models have difficulty in reproducing the strong dependence of galaxy
clustering on surface density observed in our survey.

In the framework of biased galaxy formation, our results are
consistent with a picture where more biased galaxies are more luminous
and inhabit more massive dark matter halos with a simple one-to-one
correspondence. A simple way to explain this relationship could be
that there is a direct link between the luminosity of the galaxies and
the mass of the halo. As the magnitudes we are measuring correspond to
the rest-frame ultraviolet luminosity and as we assume here there is
only one galaxy per halo, the most natural explanation of this
relationship could be a direct link between the stellar masses of the
Lyman-break galaxy population and the rest frame ultraviolet
luminosity \citep{2001ApJ...559..620P}.  However, stellar population
synthesis modelling of Lyman-break galaxies population has failed to
definitively establish such a relationship: as suggested by
\citet{2001ApJ...562...95S} these models are dependent on the assumed
extinction law, which is currently unknown for Lyman-break galaxies.

How realistic is our assumption that each Lyman-break galaxy traces
exactly one dark-matter halo? Applying semi-analytic models of galaxy
formation to the clustering of Lyman-break galaxies,
\citet{1999MNRAS.305L..21B} found that, at higher redshifts, these
models gave almost identical clustering amplitudes to these simpler
``massive halo models''. However, a more important question is how
this clustering strength scales with halo abundance. More recent work
has shown how the halo occupation function -- the number of objects
per halo -- affects sensitively the slope of the model curve in
Figure~\ref{fig:r0surf_l}
\citep{2001ApJ...554...85W,2002MNRAS.329..246B}. Models in which many
Lyman-break galaxies inhabit a single halo show a weak dependence of
clustering strength with object abundance and have difficulty
reproducing the strong trend seen in our data. 

It is also interesting to investigate the small-scale behaviour of
$\omega(\theta)$ which can provide information on the halo occupation
function \citep{2002MNRAS.329..246B}. It has been claimed that at
small ($\theta<10''$) separations $\omega(\theta)$ no longer follows a
power law \citep{2002ApJ...565...24P}. For the full $20<I_{AB}<24.5$
CFDF Lyman-break sample we have computed the ratio of pairs at small
separation $N_p(\theta<10'')$ to those at larger separation,
$N_p(10''<\theta<60'')$. Based on the fitted values of
$\omega(\theta)$ given in Table~\ref{tab:resultr0}, we expect the
ratio $N_p(10''<\theta<60'')/N_p(\theta<10'')$ to be around $19$.  In
the CFDF data (for a weighted average over all fields) we find this
pair fraction is $26\pm6$. Based on these statistics, we conclude that
the CFDF dataset provides no convincing evidence for a small-scale
departure from a power-law behaviour with $\delta=0.8$, a conclusion
consistent with the observed small-scale behaviour of $\omega(\theta)$
in Figure~\ref{fig:omtheta}.

We note that our bright measurement in Figure~\ref{fig:r0surf_l}
deviates from our model curve at the $\sim1.5\sigma$ level. We
investigate the origin of this effect, measuring the median
$(V-I)_{AB}$ colour for each of our magnitude-limited samples.  Our
brighter samples are no redder than our fainter samples, suggesting
that contamination by nearby bright ellipticals in this sample is
minimal (furthermore, all magnitude limited samples are subject to the
criterion $(V-I)_{AB}<1.0$, from Equation~\ref{eq:selbox}). A more
likely origin for this discrepancy is that in computing $r_0$, we
assume that the redshift distribution of each magnitude limited sample
is the same; a slightly lower mean redshift would imply a lower value
for $r_0$.

Finally, we remark that in our fainter bin, our stated level of
incompleteness of $\sim20\%$ at $20.0<I_{AB}<24.5$
(Table~\ref{tab:contam}) indicates that our surface densities may be
underestimated by around $\sim 0.2$. Furthermore, if Lyman-break
galaxies were especially dusty (although this is not supported by
current observations; see \citealp{webb2003}) we would expect the true
Lyman-break galaxy density to be further underestimated. However,
these considerations do not affect the principal conclusions of this
work, as these effects are expected to be much smaller than the
observed variation of clustering strength with apparent magnitude.

\subsection{Linear bias estimates for the CFDF Lyman-break sample}
\label{sec:bias}

\renewcommand{\arraystretch}{1.2}
\begin{table*}[tbp]
\begin{center}
\begin{tabular}{*{6}{c}}
{\bf Field} & magnitude & $\gamma$ & $b$ & $b$ & $b$\\
 & cuts &  & $\Omega_{0}=1.0$, & $\Omega_{0}=0.2$, &$\Omega_{0}=0.3$,\\
& ($I_{AB}$) & & $\Omega_{\Lambda}=0.0$ & $\Omega_{\Lambda}=0.0$ &$\Omega_{\Lambda}=0.7$\\
\hline               
 &  & &  &  &  \\
{\bf CFDF-14}   & 20.0--24.5 & 1.81$^{+0.21}_{-0.24}$ & 4.6$^{+5.2}_{-1.6}$ & 1.8$^{+2.0}_{-0.6}$ & 3.2$^{+3.6}_{-1.1}$ \\
          & 20.0--23.5 & 2.08$^{+0.84}_{-0.66}$ & 8.7$^{+92.6}_{-4.3}$ & 3.6$^{+38.7}_{-1.8}$ & 6.3$^{+67.5}_{-3.1}$ \\
          & 23.5--24.5 & 1.96$^{+0.25}_{-0.26}$ & 3.8$^{+5.3}_{-1.4}$ & 1.5$^{+2.1}_{-0.6}$ & 2.7$^{+3.8}_{-1.0}$ \\
{\bf CFDF-22}   & 20.0--24.5 & 1.81$^{+0.25}_{-0.35}$ & 5.4$^{+13.7}_{-2.0}$ & 2.1$^{+5.4}_{-0.8}$ & 3.7$^{+9.4}_{-1.4}$ \\
{\bf CFDF-03}   & 20.0--24.5 & 1.8 & 5.5$\pm$0.2 & 2.1$\pm$0.1 & 3.8$\pm$0.1 \\
 &  &  &  &  &  \\
{\bf CFDF mean} & 20.0--24.5 & 1.8 & 5.1$\pm$0.4 & 2.0$\pm$0.2 & {\bf 3.5$\pm$0.3} \\
{\bf CFDF-14}   & 20.0--23.5 & 1.8 & 9.9$\pm$1.5 & 3.6$\pm$0.6 & 6.4$\pm$1.0 \\
          & 23.5--24.5 & 1.8 & 4.3$\pm$0.4 & 1.7$\pm$0.2 & 3.0$\pm$0.3 \\
          &  &  &  &  &  \\  
\end{tabular}
\caption{Bias for each field and for each  magnitude
  limited sample considered in this paper, for $\overline z = 3.2$ and
  for the best fitting value of the slope.  The result marked as
  ``CFDF mean'' is computed from the mean over all three fields. The
  error bars shown correspond to Poisson error bars. To account for
  our uncertainty in the underlying redshift distribution of our
  Lyman-break sources, an extra systematic error of $\pm 0.1$ for the
  whole and faint samples and of $\pm 0.2$ for the bright sample
  should be added.}
\label{tab:resultb}
\end{center}
\end{table*}
\renewcommand{\arraystretch}{1}

The theoretical procedures described in the previous section can also
be used to estimate of the effective bias, $b$, of the Lyman-break
galaxy sample. From the comoving correlation length $r_0$ we can
compute the observed r.m.s. galaxy density fluctuation within a sphere
of $8h^{-1}$~Mpc, $\sigma_8^{gal}$ \citep{2000MNRAS.314..546M}.
Dividing this quantity by the r.m.s. mass density fluctuation, computed
from cluster-normalised models assuming the linear theory, we may
derive the linear bias $b$. In Table~\ref{tab:resultb} we present these
results, together with Poisson errors, for a range of cosmologies.

In comparison, \citet{1998ApJ...505...18A}, with a sample of
spectroscopically confirmed Lyman-break galaxies at $z\simeq3$ for
$\mathcal{R}_{AB}<25.5$, find $b=4.0\pm0.7$ for $\Omega_{0}=0.3$ and
$\Omega_{\Lambda}=0.7$.  From the average of the fainter
$I_{AB}\leq28$ galaxies selected in the HDF-N and -S,
\citet{1999MNRAS.310..540A,2002MNRAS.329..355A} find $b=1.9\pm0.4$ for
$\Omega_{0}=0.3$ and $\Omega_{\Lambda}=0.7$.

Many studies agree on the strongly biased nature of the Lyman-break
galaxy population, and provide evidence for a picture in which
structures form hierarchically and massive objects form at highest
peaks in the underlying density field
\citep{1984ApJ...284L...9K,1986ApJ...304...15B}. Our measurements of
Lyman-break galaxies at $I_{AB}=24.5$ appear to support this picture.
For very bright Lyman-break galaxies, at $I_{AB}=23.5$, we find
correlation lengths of $>10h^{-1}$~Mpc and a linear bias of $b\sim6$
in the $\Lambda$-flat cosmology. These biases would imply underlying
dark matter halo masses for the Lyman-break galaxy of around
$10^{13}h^{-1}$ M$_{\odot}$, about a factor of ten above the most
massive haloes of Lyman-break galaxy observed to date, but still
comparable to the masses of present day $M^\star$ galaxies.  We note
that the clustering lengths of our brighter Lyman-break galaxies are
comparable to those of the ``extremely red object'' (ERO) population
(e.g. $r_0=13.8\pm1.5 h^{-1}$~Mpc in a $\Lambda$-flat cosmology --
\citeauthor{2001A&A...376..825D}, \citeyear{2001A&A...376..825D}) and
we speculate that, unlike the fainter Lyman-break objects studied
previously, some fraction of these bright Lyman-break galaxies may
evolve into EROs by $z\sim1$, according to a galaxy conservation model
with a fixed bias at burst \citep{1996MNRAS.282..347M}.

\section{Summary and conclusions}
\label{sec:summary-conclusions}

We have extracted a large sample of $z\sim3$ Lyman-break galaxies from
the Canada-France Deep Fields survey.  Our catalogues cover an
effective area of $\sim 1700$ arcmin$^2$ in three separate large,
contiguous fields.  In total the survey contains 1294 Lyman-break
candidates to a limiting magnitude of $I_{AB}=24.5$.  Our conclusions
are as follows (assuming $\Omega_0=0.3$, $\Omega_{\Lambda}=0.7$):

1. Number counts and surface densities of $z\sim3$ galaxies selected in
the CFDF agrees very well with literature measurements over the entire
$20.0<I_{AB}<24.5$ magnitude range of our survey.
  
2. Using simulated catalogues, we demonstrate that at the limiting
magnitude our catalogue contains contaminants at a level of $\sim30\%$
or less.

3. We measure the two-point galaxy correlation function
$\omega(\theta)$ of Lyman-break galaxies and show it is well described
in term of a power law of slope $\delta=0.8$ even at small angular
separations, where no excess of close pairs is found.

4. Assuming that Lyman-break galaxies in the CFDF survey are at
$\overline{z}=3.2$, we derive the comoving correlation length, $r_0$,
for a range of magnitude limited samples. For the whole
$20.0<I_{AB}<24.5$ sample, we find $r_0=(5.9\pm0.5)h^{-1}$~Mpc with
the slope fixed to $\gamma=1.8$. For simultaneous fits of the slope
and amplitude , we find for the CFDF-14hr field $\gamma=1.8 \pm 0.2$
and $r_0=(5.3^{+6.8}_{-2.2})h^{-1}$~Mpc, and for the CFDF-22hr field
$\gamma = 1.8 \pm 0.3 $ and $r_0 = (6.3^{+17.9}_{-2.8})h^{-1}$~Mpc, in
good agreement with the values determined with the slope fixed.

5. In the CFDF-14hr field, we find a marginal dependence of $r_0$ on
apparent magnitude: for Lyman-break galaxies with $20.0<I_{AB}<23.5$,
we derive $r_0=(11.6\pm2.0)h^{-1}$~Mpc, whereas for $23.5<I_{AB}<24.5$
we find $r_0=(5.0\pm0.6)h^{-1}$~Mpc (in both cases for slopes fixed to
$\gamma=1.8$).  Allowing both the slope and amplitude to vary, this
segregation is still detected at the $3\sigma$-level.

6. We investigate the dependence of $r_0$ on surface density, $n$, and
find a strong correlation. For $n=(0.09\pm0.02)$~arcmin$^{-2}$,
$r_0=(11.6\pm2.0)h^{-1}$~Mpc, whereas for
$n=(0.78\pm0.24)$~arcmin$^{-2}$, we find $r_0=(5.9\pm0.5)h^{-1}$~Mpc.

7. A simple analytic model in which each Lyman-break galaxy traces one
dark matter halo is able to reproduce the observed dependence of
correlation length on abundance quite well, except for our very
bright sample of Lyman-break galaxies, which deviates from the
predictions of our models by around $\sim 1.5\sigma$.

8. We derived a linear bias $b$ by dividing the measured r.m.s.  galaxy
density fluctuation $\sigma_8^{gal}$ by the r.m.s. mass fluctuation
$\sigma_8^{m}$ computed by assuming cluster-normalised linear theory.
For our sample of Lyman-break galaxies, we find for $20.0<I_{AB}<24.5$,
$b=3.5\pm0.3$.

\begin{acknowledgements}
  SF and HJMCC wish to acknowledge the use of TERAPIX computer
  facilities at the Institut d'Astrophysique de Paris, where part of
  the work for this paper was carried out. We would like to thank
  Chuck Steidel for providing us with his Lyman-break catalogue
  covering the 14hr field. We would also like to thank our referee for
  a detailed and thorough report which improved our paper.  HJMCC's
  work has been supported by MIUR postdoctoral grant COFIN-00-02 and a
  VIRMOS postdoctoral fellowship.
\end{acknowledgements}

\bibliographystyle{aa}

\end{document}